\documentclass[letterpaper,groupedaddress]{elsart}
\usepackage{graphicx,amssymb,amsmath,times,wrapfig,units,fancyhdr}
\usepackage{geometry}
\journal{Astroparticle Physics}
\newif\ifbw\bwfalse
\begin{document}
\begin{frontmatter}
\date{\today}
\title{Observation in the
  MINOS far detector of the shadowing of cosmic rays by the sun and moon}

\author[FNAL]{P.~Adamson} 
\author[RAL,Athens]{C.~Andreopoulos} 
\author[ANL]{D.~S.~Ayres} 
\author[Oxford]{C.~Backhouse} 
\author[Oxford]{G.~Barr} 
\author[Washington]{W.~L.~Barrett} 
\author[BNL]{M.~Bishai} 
\author[Cambridge]{A.~Blake} 
\author[Duluth]{B.~Bock} 
\author[FNAL]{G.~J.~Bock} 
\author[FNAL]{D.~J.~Boehnlein} 
\author[FNAL]{D.~Bogert} 
\author[Indiana]{C.~Bower} 
\author[ANL]{S.~Budd}
\author[Harvard]{S.~Cavanaugh} 
\author[Tufts]{D.~Cherdack} 
\author[FNAL]{S.~Childress} 
\author[FNAL,Caltech]{B.~C.~Choudhary\thanksref{Delhi}} 
\author[Oxford]{J.~H.~Cobb} 
\author[UNICAMP]{J.~A.~B.~Coelho} 
\author[WandM]{S.~J.~Coleman} 
\author[Indiana]{L.~Corwin} 
\author[Minnesota]{D.~Cronin-Hennessy} 
\author[Pittsburgh]{I.~Z.~Danko} 
\author[Oxford,IIT]{J.~K.~de~Jong} 
\author[Sussex]{N.~E.~Devenish} 
\author[BNL]{M.~V.~Diwan} 
\author[UCL]{M.~Dorman} 
\author[UNICAMP]{C.~O.~Escobar} 
\author[UCL]{J.~J.~Evans} 
\author[Sussex]{E.~Falk} 
\author[Harvard]{G.~J.~Feldman} 
\author[HolyCross,Benedictine]{M.~V.~Frohne} 
\author[Tufts]{H.~R.~Gallagher} 
\author[Carolina]{A.~Godley} 
\author[Goias]{R.~Gomes}
\author[ANL]{M.~C.~Goodman} 
\author[USP]{P.~Gouffon} 
\author[Duluth]{R.~Gran} 
\author[RAL]{N.~Grant}
\author[Minnesota,Duluth]{E.~W.~Grashorn\corauthref{cor}\thanksref{Osu}}
% ...

\thanks[Alabama]{Now at Department of Physics and Astronomy, University of Alabama, Tuscaloosa, Alabama 35487, USA.}
\address[ANL]{Argonne National Laboratory, Argonne, Illinois 60439, USA} 
\address[Athens]{Department of Physics, University of Athens, GR-15771 Athens, Greece} 
\address[Benedictine]{Physics Department, Benedictine University, Lisle, Illinois 60532, USA} 
\address[BNL]{Brookhaven National Laboratory, Upton, New York 11973, USA} 
\address[Caltech]{Lauritsen Laboratory, California Institute of Technology, Pasadena, California 91125, USA} 
\address[Cambridge]{Cavendish Laboratory, University of Cambridge, Madingley Road, Cambridge CB3 0HE, United Kingdom} 
\thanks[Delhi]{Now at Department of Physics and Astrophysics, University of Delhi, Delhi 110007, India.} 
\address[FNAL]{Fermi National Accelerator Laboratory, Batavia, Illinois 60510, USA} 
\address[Goias]{Universidade Federal de Goias, Instituto de Fisica, CP 131, 74001-970, Goiania, Goias, Brazil}
\address[Harvard]{Department of Physics, Harvard University, Cambridge, Massachusetts 02138, USA} 
\address[HolyCross]{Holy Cross College, Notre Dame, Indiana 46556, USA} 
\address[IIT]{Physics Division, Illinois Institute of Technology, Chicago, Illinois 60616, USA} 
\address[Iowa]{Department of Physics and Astronomy, Iowa State University, Ames, Iowa 50011, USA} 
\address[Indiana]{Indiana University, Bloomington, Indiana 47405, USA} 
\thanks[LASL]{Now at Nuclear Nonproliferation Division, Threat Reduction Directorate, Los Alamos National Laboratory, Los Alamos,
New Mexico 87545, USA.} 
\address[Lebedev]{Nuclear Physics Department, Lebedev Physical Institute, Leninsky Prospect 53, 119991 Moscow, Russia} 
\address[LLL]{Lawrence Livermore National Laboratory, Livermore, California 94550, USA} 
\address[UCL]{Department of Physics and Astronomy, University College London, Gower Street, London WC1E 6BT, United Kingdom} 
\address[Minnesota]{University of Minnesota, Minneapolis, Minnesota 55455, USA} 
\address[Duluth]{Department of Physics, University of Minnesota -- Duluth, Duluth, Minnesota 55812, USA} 
\thanks[Osu]{Now at Department of Physics, Ohio State University, Columbus, Ohio 42310, USA.} 
\address[Otterbein]{Otterbein College, Westerville, Ohio 43081, USA} 
\address[Oxford]{Subdepartment of Particle Physics, University of Oxford, Oxford OX1 3RH, United Kingdom} 
\address[Pittsburgh]{Department of Physics and Astronomy, University of Pittsburgh, Pittsburgh, Pennsylvania 15260, USA} 
\address[IHEP]{Institute for High Energy Physics, Protvino, Moscow Region RU-140284, Russia} 
\address[Carolina]{Department of Physics and Astronomy, University of South Carolina, Columbia, South Carolina 29208, USA} 
\address[RAL]{Rutherford Appleton Laboratory, Science and Technology Facilities Council, OX11 0QX, United Kingdom} 
\address[Stanford]{Department of Physics, Stanford University, Stanford, California 94305, USA} 
\thanks[StJohnFisher]{Now at Physics Department, St. John Fisher College, Rochester, New York 14618 USA.} 
\address[Sussex]{Department of Physics and Astronomy, University of Sussex, Falmer, Brighton BN1 9QH, United Kingdom} 
\address[TexasAM]{Physics Department, Texas A\&M University, College Station, Texas 77843, USA} 
\address[Texas]{Department of Physics, University of Texas at Austin, 1 University Station C1600, Austin, Texas 78712, USA} 
\address[Tufts]{Physics Department, Tufts University, Medford, Massachusetts 02155, USA} 
\address[UNICAMP]{Universidade Estadual de Campinas, IFGW-UNICAMP, CP 6165, 13083-970, Campinas, SP, Brazil} 
\address[USP]{Instituto de F\'{i}sica, Universidade de S\~{a}o Paulo,  CP 66318, 05315-970, S\~{a}o Paulo, SP, Brazil} 
\address[Warsaw]{Department of Physics, University of Warsaw, Ho\.{z}a 69, PL-00-681 Warsaw, Poland} 
\address[Washington]{Physics Department, Western Washington University, Bellingham, Washington 98225, USA} 
\address[WandM]{Department of Physics, College of William \& Mary, Williamsburg, Virginia 23187, USA} 
\thanks[deceased]{Deceased.}

\begin{abstract} 
The shadowing of cosmic ray primaries by the 
the moon and sun
  was observed  by the MINOS far detector at a depth of
  \unit[2070]{mwe} using 83.54 million cosmic ray muons accumulated over
  1857.91 live-days.  The shadow of the moon was detected at the
  \unit[5.6]{$\sigma$} level and the shadow of the sun at the
  \unit[3.8]{$\sigma$} level using a log-likelihood search in celestial
  coordinates.  The moon shadow was used to
  quantify the absolute astrophysical pointing of the detector to be
  0.17$\pm$0.12$^\circ$.  
Hints of Interplanetary
Magnetic Field effects were observed in both the sun and moon shadow.  

\end{abstract}
\end{frontmatter} 

\section{Introduction \& Motivation}\label{sec:intro}
The Main Injector Neutrino Oscillation Search (MINOS) far detector~\cite{MinosNIM} is a magnetized scintillator and steel
tracking calorimeter, located in the Soudan Mine, ( $47^{\circ}$
49' 13.3'' N, $92^{\circ}$ 14' 28.5'' W)  in 
northern Minnesota, USA, at a depth of 2070 meters water
equivalent (mwe). 
While the primary function of the far detector is to detect
neutrinos from Fermilab's NuMI $\nu_{\mu}$ beam~\cite{Adamson:2007gu}, the great depth and wide
acceptance of the detector combined with the flat overburden of the
Soudan site allow it to serve as an efficient cosmic-ray muon detector.
The \unit[5.4]{kton} detector is composed of 486~\unit[8]{m} wide octagonal
planes, each consisting of a \unit[2.54]{cm} thick steel plate and a \unit[1.0]{cm} thick scintillator
plane.  Adjacent planes are separated by a \unit[2.4]{cm} air gap.  It  is \unit[30]{m} long and has a total
aperture of  \unit[6.91$\times 10^6$]{cm$^2$
sr}~\cite{Rebel:2004mm} for the cosmic rays selected in this analysis.
MINOS observes underground muons with a minimum surface energy of
\unit[0.7]{TeV}, and the sharply peaked energy spectrum has a mean value
of about \unit[1.0]{TeV}~\cite{Adamson:2009zf}.  This mean muon energy
corresponds to a mean cosmic ray
primary energy of about \unit[10]{TeV}.
  
Optical telescopes use a standard catalog of stars to establish the
resolution and pointing reliability of a new instrument.  This is not
possible for a cosmic ray detector as there are no known cosmic ray sources
available for calibration~\cite{Ambrosio:2002wr}.  
The shadow caused by the absorption of cosmic rays by the moon is a
well observed phenomenon in the otherwise isotropic cosmic ray sky~\cite{Clark:1957}. 
This shadow
provides a means of studying the resolution and
alignment of the detector which are important in the search for cosmic
point sources.  The physical extent and shape of the 
shadow gives information about the resolution of the detector, while
the location of the deficit center measures the absolute
pointing of the detector.   The moon has a 0.52$^{\circ}$ diameter as viewed
from earth, and the cosmic ray deficit it causes has
been measured by air shower arrays (CYGNUS~\cite{Alexandreas:1990wj},
CASA~\cite{Borione:1993xq}, Tibet~\cite{Amenomori:1993iv},
Milagro~\cite{Atkins:1999gb}, GRAPES~\cite{Grapes}, HEGRA~\cite{Hegra}) as well 
as underground detectors (Soudan~2~\cite{Cobb:1999mi},
MACRO~\cite{Ambrosio:1998wv,Ambrosio:2003mz}, L3+C~\cite{Achard:2005az},
BUST~\cite{Bust}).  

The shadow of the moon is affected by multiple Coulomb
scattering, the geomagnetic field and the Interplanetary Magnetic Field
(IMF)~\cite{Cobb:1999mi}.  Multiple Coulomb scattering occurs 
in the rock overburden and  
causes a general spreading  of the moon deficit disc.  The geomagnetic
field is nearly a dipole and causes an eastward deflection of positive
primaries, which results 
in a positive horizontal shift in the observed shadow of up to $\Delta
\alpha$=0.15$^{\circ}Z/E_{p}$(TeV)~(see Sec.~\ref{sub:geomag}).  Older
calculations using an impulse approximation give a larger value~\cite{Urban:1989eu,Heintze:1990mq}.  
The IMF is produced by the sun, which has an ambient dipole field that is
100 times greater than the geomagnetic field.  The field is carried
through the solar system by the solar wind, the stream of energetic charged
particles that emanate from the atmosphere of the sun.  Since the sun
has a 27 day rotation period, the magnetic field has a spiral shape, called a Parker
spiral~\cite{Parker:1958}.  The IMF causes a deflection of primaries that
strongly depends on the solar wind.  It changes in time and has a 
sectorized structure~\cite{Wilcox:1965}.  This complex structure makes
it hard to model. The direction of the IMF vector is either ``towards''
or ``away'' from the earth and the two field directions are separated by
a thin, field-free, region known as the ``neutral current sheet''. 
 The IMF causes a deflection that smears the moon's shadow,
though this effect is small since a cosmic ray primary travels a
relatively short distance from the moon to earth.  The larger IMF effect
comes from its distortion of the geomagnetic field.
 
The sun also subtends a disk of diameter $ 0.52^{\circ}$, and in principle produces a similar shadow to the moon. Its shadow has been observed by
CYGNUS~\cite{Alexandreas:1990wj},
Tibet~\cite{Amenomori:1993iv,Amenomori:1993xj} and
MACRO~\cite{Ambrosio:2003mz}.
However, the much greater distance that cosmic primaries travel in the IMF can produce significant distortions in the shadow.
The IMF varies according to the 11 year solar activity cycle and peaks at solar maximum, when the
sun's magnetic field changes polarity~\cite{SolarCycle}.   The most recent solar maximum
occurred in February, 2001, and the following  minimum occurred in
December, 2008.  The average magnitude of the IMF over the period of time that the MINOS
data were collected was \unit[6.5]{nT}.  The maximum value was
\unit[50]{nT}. Assuming a constant, uniform field and a \unit[10]{TeV}
proton,  the average magnetic deflection is about 0.05$^{\circ}$.  For
 a \unit[10]{TeV} proton traveling along the line from the Sun to the earth, the
deflection is up to 0.75$^{\circ}$.  Because of the complex
structure of the IMF already mentioned, the assumption of a constant,
uniform field is not correct.  By measuring the behavior of an
ensemble of particles, thus sampling many parts of the IMF, a more refined
understanding of the average IMF properties can be obtained.  

\subsection{Spherical Coordinate Systems}
Horizon coordinates use the detector's local horizon as the fundamental
plane.  The longitudinal angle is azimuth ($Az$), measured east from north
and ranges from $0^{\circ}$ to $360^{\circ}$.  The angle out of the equatorial
plane is zenith ($Zen$), where vertical up has a value of  $0^{\circ}$ and
the horizon has a value of  $90^{\circ}$.  Zenith is measured from
$0^{\circ}$ to $180^{\circ}$. 
Celestial coordinates are projected on the sky and centered on the earth's
equatorial plane.  The longitudinal
angle is right ascension ($RA$), measured eastward from the
Vernal equinox and ranges from $0^{\circ}$ to $360^{\circ}$. 
The angle out of the equatorial plane is declination ($Dec$), ranging between $\pm90^{\circ}$.  
Ecliptic coordinates are celestial coordinates that use the ecliptic
(the path the sun follows over the course of a year) as its
fundamental plane.  
The longitudinal 
angle is ecliptic longitude ($\lambda$), and is the same as $RA$. 
The angle out of the plane of the ecliptic is ecliptic latitude
($\beta$), ranging between $\pm90^{\circ}$. 

\section{Data}\label{sec:data}
In order to perform the shadowing analysis, the muon data must be
selected for reliable pointing, the smearing and systematic offsets
of the shadow must be understood, and the backgrounds must be
quantified.  

\subsection{Event Selection}\label{sec:event_sel}
 This analysis encompassed events recorded over 1980 days, from August~1,~2003 to December~31,~2008, giving a total of
1857.91 live-days when the detector was operational.  The data set includes 83.54 million cosmic ray induced muon
tracks.  Cosmic~ray muons were triggered by recording hits on four
planes within a group of five planes. 
Several selection criteria were
required to ensure that the detector was in a reliable state when the
data were taken (Pre-Analysis selection criteria) and that only well reconstructed tracks
were included in the sample (Analysis selection criteria).  The Pre-Analysis selection criteria are
described in~\cite{Adamson:2007ww}.
%
The following analysis selection criteriawere applied:
\begin{enumerate}
\item ``Number of Planes $>$~9'', a track that passes fewer planes may
      not give reliable localization information to the track fitter. 
\item ``Track Length $>$~\unit[1.55]{m}'', any event with a track shorter than \unit[1.55]{m}
      may not be reliably reconstructed.   
\item ``$|
      \overrightarrow{\sigma}_{vtx}-\overrightarrow{\sigma}_{end}|<$~\unit[0.021]{m}''
      If the uncertainty on the endpoint position, $\overrightarrow{\sigma}_{end}$, 
      is significantly different from the uncertainty on the beginning
      position, $\overrightarrow{\sigma}_{vtx}$, then the muon has  
      questionable reconstruction pointing. $|\overrightarrow{\sigma}_{vtx}-\overrightarrow{\sigma}_{end}|$ is a
      measure of the absolute value of the difference in track beginning and endpoint uncertainty.
\end{enumerate}
These selection criteriawere chosen to optimize the selection of cosmic ray
muons with good pointing resolution on the sky.
The cut values were determined empirically using standard MINOS \textsl{Monte
  Carlo} events, inserting the moon, and maximizing the moon shadow. 
The number of muons that survived each cut is shown in
Table~\ref{tab:cuttable}. Of the initial 83.54 million 
triggers, 62.5\% survived all cuts, leaving 52.19 million muons.

\begin{table}[h]
\centering
\caption {Fraction of events that survive each
pointing cut}
\begin{tabular}{||l|c|c||} \hline\hline
\label{tab:cuttable}
\textbf{Cut}  & \textbf{No. Remaining} &\textbf{Fraction Remaining} \\\hline\hline
 Total Tracks & 83.54$\times10^6$ & 1.0\\\hline 
\hspace*{0.5cm}1. Data Quality Cuts~\cite{Adamson:2007ww} & 68.91$\times10^6$ & 0.825 \\\hline \hline
\hspace*{0.5cm}2. Number of Planes $<$ 10 & 62.06$\times10^6$ & 0.743\\\hline 
\hspace*{0.5cm}3. Track Length $<$\unit[1.55]{m} & 60.10$\times10^6$ & 0.730 \\\hline 
\hspace*{0.5cm}4. $|\overrightarrow{\sigma}_{vtx}-\overrightarrow{\sigma}_{end}|
> $0.021 & 52.19$\times10^6$ & 0.625 \\\hline 
\end{tabular}
\end{table}

There are 17,389 muons in a 2$^{\circ}$ half-angle cone centered on the moon and
16,411 muons in a 2$^{\circ}$ half-angle cone centered on the sun.  The reason for
this difference is that the detector is at a high latitude, thus the number of 
muons collected in winter near the sun's location will be fewer than the
number collected near the moon. 
The data set includes five full yearly cycles plus
 five months.  Those extra months came after the summer
solstice, so the amount of time the sun spends above the horizon (and
the angle of the sun above the horizon)
continues to decrease as the period progresses.

\subsection{Geomagnetic Field}\label{sub:geomag}
\begin{figure}[h!]\begin{center}
\ifbw
\includegraphics[width=0.8\textwidth]{GeomagSepMax-BW}
\else 
\includegraphics[width=0.8\textwidth]{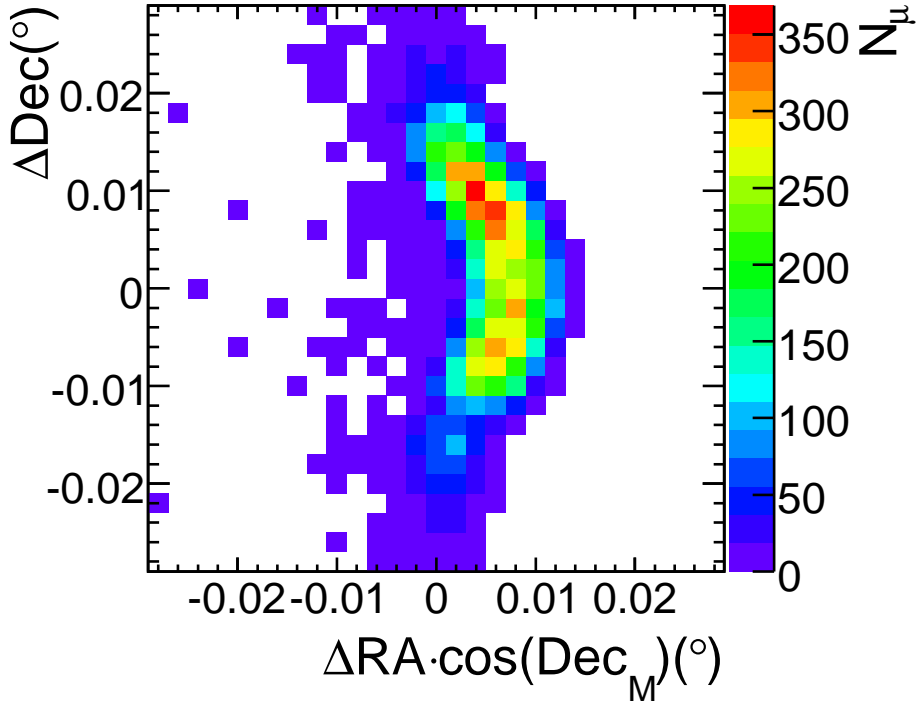}
\fi
\vspace{-10pt}
\caption[Mean Geomagnetic Deflection]{\label{fig:geomean} The distribution
  of integrated geomagnetic deflections for cosmic rays that pass near
  the moon in horizon coordinates.  This distribution is for the case where the muon
  energy was determined from the far detector overburden profile and each primary cosmic
  proton had energy $10p_{\mu}$.} 
\end{center}
\end{figure}

The earth's magnetic field produces a relative east-west shift in the apparent
arrival direction of cosmic ray primaries from the direction of the moon.
Older experiments~(\cite{Cobb:1999mi,Ambrosio:1998wv,Ambrosio:2003mz}) 
used  the ``impulse approximation'' to calculate
the expected \unit[10]{TeV} proton deflection, 
0.15$^{\circ}$.  
Since the path a particle travels is fairly long, the 
``impulse approximation'' may not be valid over such large distances.
The geomagnetic deflection $\Delta \alpha$ 
of a given cosmic ray with charge $Z$ and momentum $p$(TeV/c), depends strongly on the local magnetic field
conditions, $\overrightarrow{B}$
and the particle path
$\overrightarrow{l}$.
It can be written:
\begin{equation}\label{eq:geomag_eq}
\tan(\Delta\alpha) = \frac{Z}{p}\int_0^{d_{m}}
\overrightarrow{B}  
\times d \overrightarrow{l},
\end{equation}
where \unit[$d_m=384\times10^3$]{km} is the distance from the earth to
the moon.  To
measure the effect of geomagnetic deflection on the moon shadow
the magnetic field
encountered by a cosmic ray primary between the moon and the earth was
integrated along the cosmic ray path.  The integral was calculated
numerically, with the cross product computed at \unit[20]{km} intervals.
Muons with arrival
directions within a $2^{\circ}$ half-angle cone of
the moon's location were chosen from the data set.  The geomagnetic field from the surface of the
earth to \unit[600]{km} was found using
the International Geomagnetic Reference Field (IGRF)
calculation~\cite{IGRF}. The geomagnetic field from \unit[600]{km} to
the magnetopause 
(which varies depending on the IMF, here taken to be \unit[70,000]{km})  
was found using GEOPACK~\cite{Tsyganenko:1995,Tsyganenko:1996}.  
The geomagnetic field from the magnetopause to the location of the moon
(\unit[384,000]{km}) was held constant at the value at \unit[70,000]{km}.  The muons that survive to the depth of the far detector are usually the decay products of
mesons created in the first cosmic ray interaction, typically at a height of
\unit[20]{km}.  The integrals were thus calculated using  $\mu^+$ with
momentum $p_{\mu} = 0.1p_{p}$ from
\unit[0-20]{km} and protons from
\unit[20-384,000]{km}.
The result of this calculation can be seen in Fig.~\ref{fig:geomean}, where the
surface energy of each muon was set to \unit[0.7]{TeV}, the vertical
muon threshold energy of the far detector, and that $p_p=10p_{\mu}$.
The maximum deflection was $\Delta RA = 0.02^{\circ}$, and the mean
deflection was $\Delta RA = 0.005^{\circ}$.   
Note that while
there was a clear positive shift in $RA$, the distribution of deflections
in $Dec$ was more uniform.  Any displacement of the moon shadow beyond $0.015^{\circ}$ cannot be
caused by the geomagnetic field, and it is unlikely that the geomagnetic
field could cause an average deflection of greater than $0.01^{\circ}$.  

\subsection{Multiple Muons}\label{sub:dimuons}
\begin{figure}[h!]
\begin{center}
\includegraphics[width=1.0\textwidth]{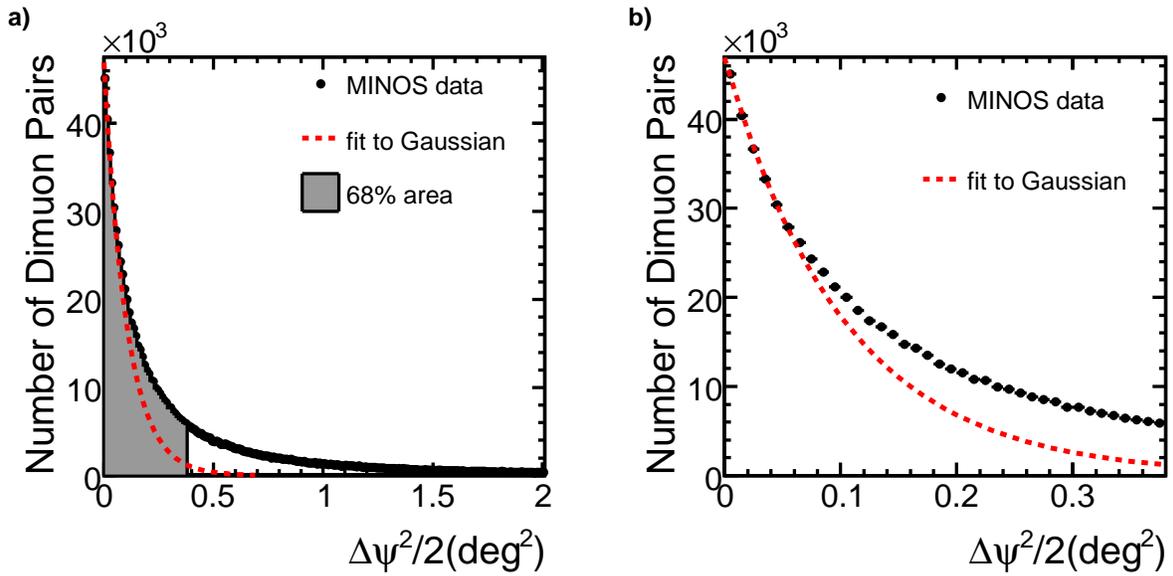}
\vspace{-10pt}
\caption[Multiple Muon Separation]{\label{fig:resolution} a) The 
square of the single muon
angular separation of multiple muon
pairs, $psi$, in deg$^2$.  The shaded region represents 68\% of the 
distribution, which is
taken to be the resolution of the detector,
$\psi/\sqrt{2}=0.62^{\circ}$.  The peak 
region of the distribution was fitted with a Gaussian function, shown by
the dashed line.  The fit parameter $\sigma=0.322\pm0.002^{\circ}$. b) The 
square of the single muon angular separation of multiple muon
pairs out to 0.375~deg$^2$. } 
\end{center}
\end{figure}

In addition to the smearing effects of the
geomagnetic and interplanetary magnetic fields, the cosmic muon angular
resolution of an underground detector is limited by multiple Coulomb 
scattering in the surrounding rock as well as the geometric resolution of the
apparatus itself.  The geometric resolution of the detector is of order
$\pm 0.15^{\circ}$ but the dominant effect \unit[2070]{mwe} underground is
multiple Coulomb scattering.  
The angular resolution of
the detector was measured using 3.12 million multiple muon events with
reconstructed multiplicity equal to two, collected from
August~1,~2003  to December~31,~2007.  
Multiple muon events are created by a cosmic ray of sufficient energy
to generate more than one energetic muon.  The energy spectrum of
multiple muon is similar to that of single muons, which is a steeply falling
power law with a spectral index approximately minus three.  Most muons (multiple muon
events included) are close to the low energy cutoff.   The transverse momentum of
these pairs 
is negligible compared to their longitudinal momentum, and they have
angular separation $<$0.05$^{\circ}$~\cite{Ambrosio:1998wv} at creation.
The measured angular separation between these pairs quantifies the resolution of the detector.  
Pairs were selected by the same criteria used for single muons (see
Sec.~\ref{sec:event_sel}).  A total of 1.77 million multiple muon events
survived. The distribution
of their angular separation $\psi$, divided by $\sqrt{2}$ to apply to the single
muon resolution, is shown in Fig.~\ref{fig:resolution}.  
The resolution is defined to be the single muon angular separation within
which 68\% of the 
distribution lies.
This value, shown by the shaded region in Fig.~\ref{fig:resolution}, is 0.62$^{\circ}$.  The peak
region of the distribution was fitted with a Gaussian function, shown by the dashed
line.  The fit parameter $\sigma=0.322\pm0.002$.  The Gaussian function
fits well over the first few bins, which describes the bulk of the distribution, but the
long tail begins to deviate from the Gaussian function
at $\psi^2/2 \sim \unit[0.06]{deg^2}$.  The long tail can be attributed to
Moliere scattering in the rock overburden~\cite{Cobb:1999mi}.

\subsection{\textsl{Monte Carlo} Simulation}\label{sub:bkgnd}
The backgrounds for the shadow analyses were calculated using a
\textsl{Monte Carlo} simulation. A muon arrival direction was chosen out of the known distribution of
events in the detector (in horizon coordinates) and paired with a random
time chosen from the 
known time distribution to find the muon's location in celestial
coordinates.  This was done for every data muon to create
one background sample, and 1,000 background samples were created.

\begin{figure}[!h]
\begin{center}
\begin{minipage}[l]{0.50\linewidth}
\ifbw
\includegraphics[width=1.0\textwidth]{MoonTempl-BW}
\else
\includegraphics[width=1.0\textwidth]{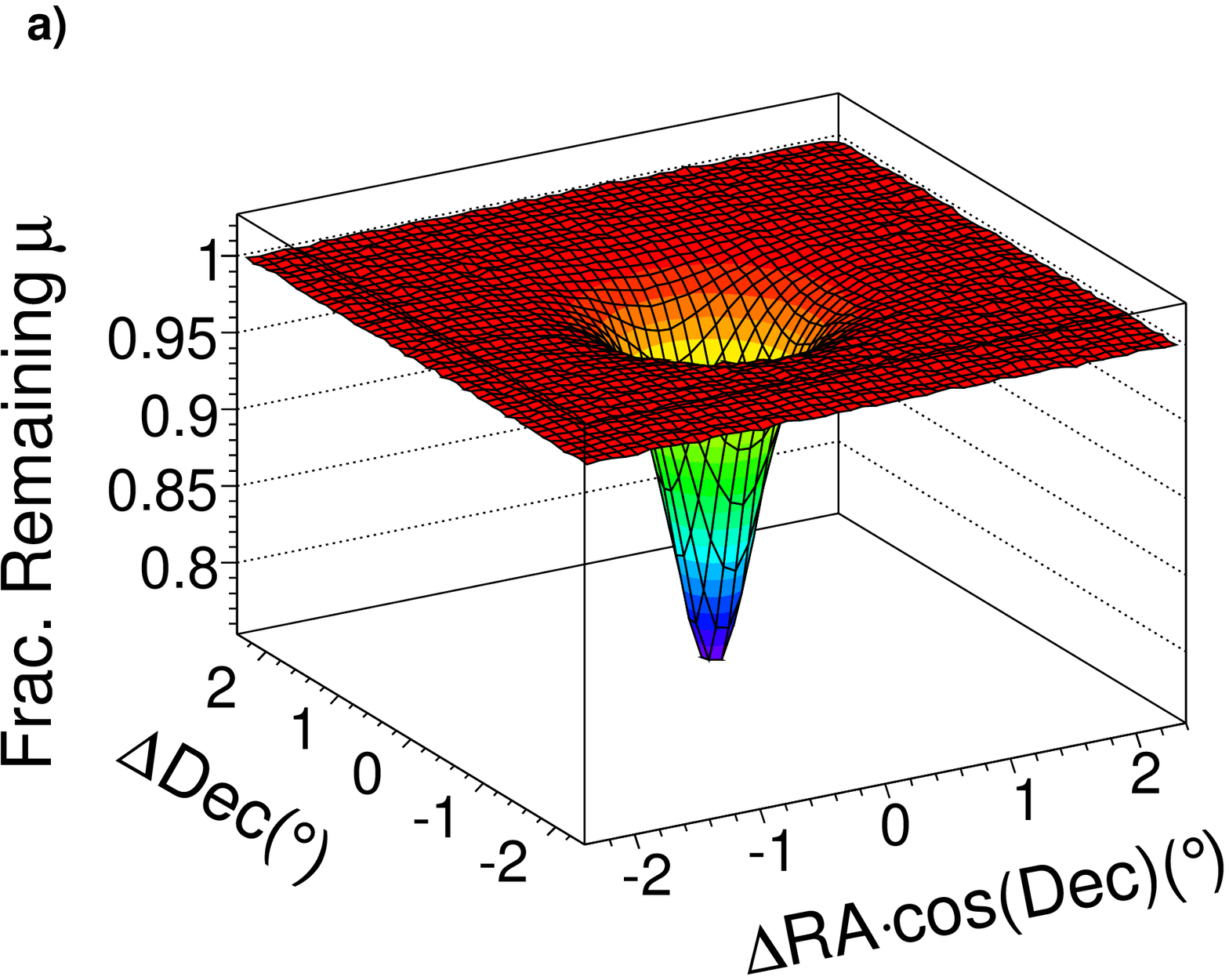}
\fi
\end{minipage}
\begin{minipage}[r]{0.49\linewidth}
\ifbw
\includegraphics[width=0.99\textwidth]{MoonTempl1-BW}
\else
\includegraphics[width=0.99\textwidth]{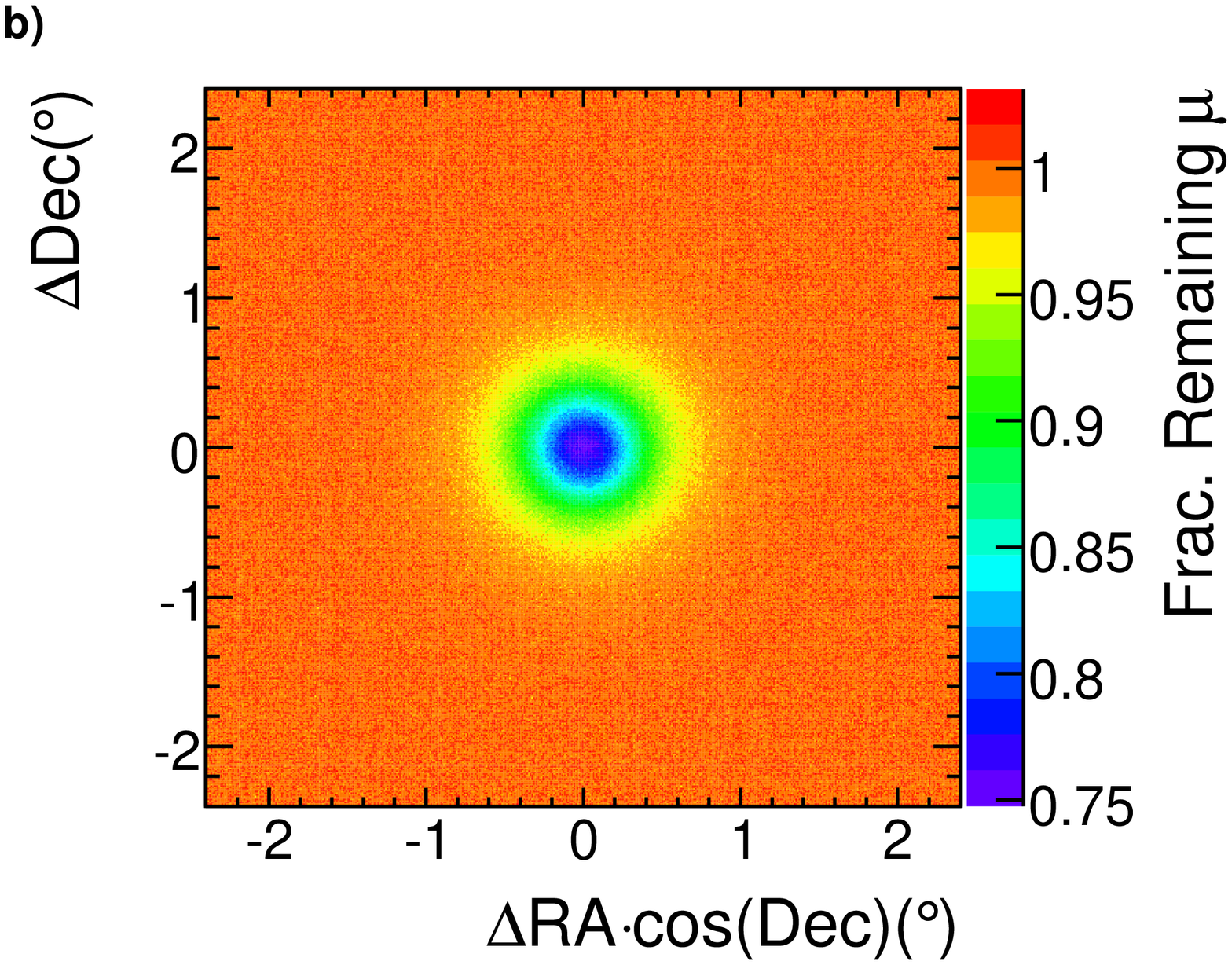}
\fi
\end{minipage}
\caption[Moon Template]{\label{fig:ExpectedShadow} The expected cosmic 
ray shadow of 
the moon (from simulation) as seen in the far detector, in two
different views: a) in three dimensions, and b) in a two dimensional
projection of a).  It is shown in celestial coordinates centered on the
location of the moon.  The color axis shows the surviving fraction of cosmic-ray
induced muons.  } 
\end{center}
\end{figure}

A template of the expected distribution of the sun and moon shadows in
celestial coordinates (right ascension ($RA$) and declination ($Dec$)) was
simulated using the detector resolution determined from the multiple muon
events in Sec.~\ref{sub:dimuons} to account for the smearing effects of
coulomb scattering and detector resolution.  Cosmic rays were generated
traveling towards a disk the size of the moon or sun.  If the cosmic ray
intercepted the disk it was removed.  If not, the cosmic ray was assumed
to produce a muon that would travel underground to the detector, and an angular deviation was
selected at random from the multiple muon distribution.  The resulting
distribution in $\Delta RA\cdot\cos(Dec)$ ($\cos(Dec)$ normalizes for
solid angle)  and $\Delta Dec$ can be seen for the moon in
Fig.~\ref{fig:ExpectedShadow}. 

\section{Moon Shadow}\label{sec:shadowing}
\subsection{Two Dimensional Shadow}\label{sub:2dms}
 A two dimensional maximum likelihood grid search was used to find the
 location of the moon.
The muons in the background samples generated
in Sec.~\ref{sub:bkgnd} were sorted by separation from the moon's
location in equal solid
angle bins $0.01^{\circ}$ on a side.  The x coordinate is given by
 $\Delta RA = (RA_{\mu}-RA_M)\cdot \cos(Dec_M)$, where the subscripts
 $\mu$ and $M$ denote muon and moon, respectively, and the y coordinate is
 given by   $\Delta Dec = (Dec_{\mu}-Dec_M)$.  $\Delta RA$ is modified
 by $\cos(Dec_M)$ to account for the projection of sphere onto a flat grid. 
One thousand cosmic ray background samples were averaged
to create a smooth, isotropic background grid.  The data were sorted in a
similar grid.
A grid search utilizing a log-likelihood method was employed to find the
most probable position of a moon-like deficit.  This method was invented
by Cash~\cite{Cash:1979vz}, first applied by COS-B~\cite{CosB}, and
first applied to the moon shadow by MACRO~\cite{Ambrosio:1998wv}.  The
moon shadow template 
(Sec.~\ref{sub:bkgnd}) was 
placed at a fixed position $(x_s,y_s)$ on the data grid.  The search was
then performed by integrating over the the shadow template at this
location.  The template was then moved to adjacent bin location
$(x_s,y_{s+1}$), and the search was repeated.  This process was repeated until the entire 4$^{\circ}$
by 4$^{\circ}$ data grid was scanned.  The shadow
that fit the data best was found by maximizing the shadow strength
$I_s$ using the likelihood function: 
\begin{equation}\label{eq:loglikely}
l(x,y,I_{s}) = 2 \sum_{i=1}^{n_{bin}}\left[N_{i}^{th}-N_{i}^{obs}+
N_{i}^{obs} \ln \frac{ N_{i}^{obs}}{N_{i}^{th}}\right],
\end{equation}
where $N_{i}^{th} = N_{i}^{back}-I_s \cdot P_s(x_i,y_i)$ is the number
of events expected in bin $i$, $N_{i}^{back}$ is the number
of muons from the smoothed background grid in bin $i$ and $P_s(x_i,y_i)$ is the
fraction of the cosmic rays at location $(x_i,y_i)$ blocked by the
moon, and is equal to one minus the distribution in
Fig~\ref{fig:ExpectedShadow}. $I_s \cdot P_s(x_i,y_i)$ is the number of
events removed from bin $i$ 
by the moon.  To determine the
strength of this deficit, the parameter $\Lambda$ was defined as:
\begin{equation}\label{eq:labmdadef}
\Lambda = l(x,y,0) - l(x,y,I_s),
\end{equation}
which is a measure of the deviation from the null (no-moon) hypothesis.
 
\begin{figure}[h!]
\begin{center}
\ifbw
\includegraphics[width=0.8\textwidth]{2DMSAll-BW}
\else
\includegraphics[width=0.8\textwidth]{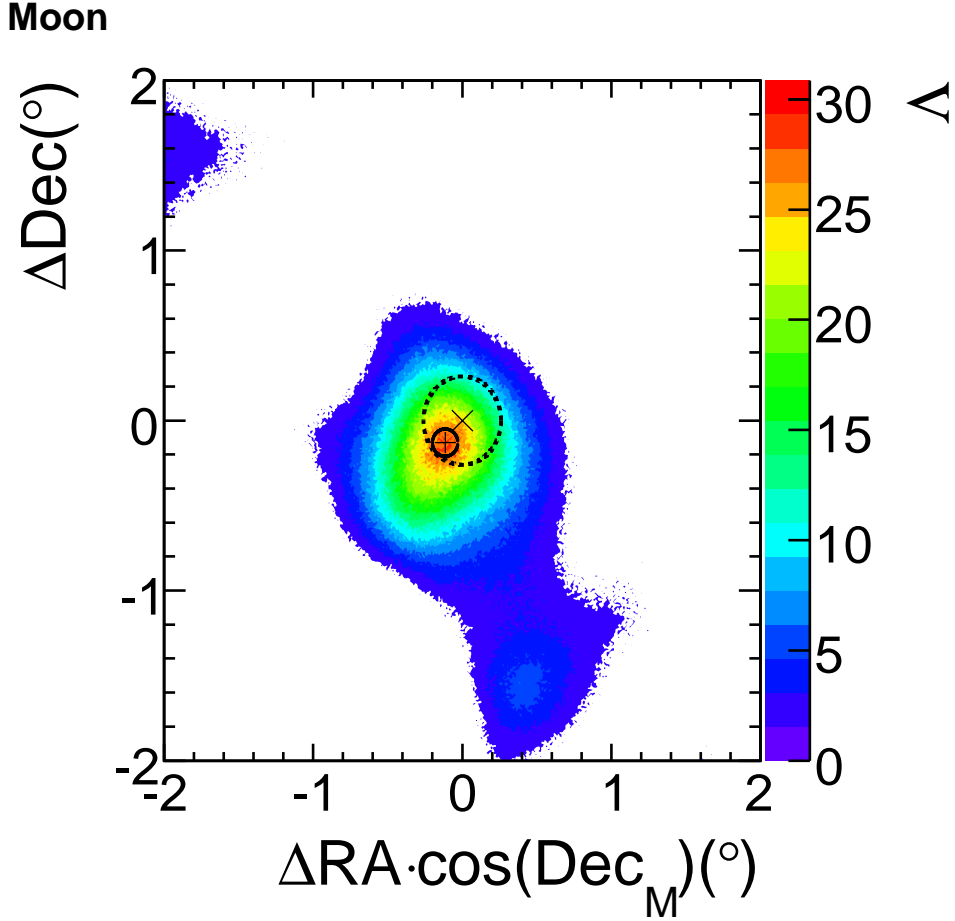}
\fi
\vspace{-10pt}
\caption[Two Dimensional Moonshadow]{\label{fig:2DMShadow} The two dimensional moon induced muon
deficit in $\unit[0.01]{^{\circ}}$ wide bins, in celestial coordinates.  The $\times$
marks the expected location of the moon, and the dashed circle is the apparent
size of the moon as viewed from earth.   The cross marks the observed
location of the moon, and the solid ellipse denotes the uncertainty.  The greatest deficit is $\Lambda =
30.9$, a \unit[5.6]{$\sigma$} significance, centered on $(-0.11^{\circ},-0.13^{\circ})$ } 
\end{center}
\end{figure}

The two dimensional distribution of these deviations was drawn on a
$4^{\circ} \times 4^{\circ}$ grid, binned in $0.01^{\circ}$ on a side,
and can be seen in
Fig.~\ref{fig:2DMShadow} in celestial coordinates.
The greatest deficit 
is $\Lambda = 30.9$, centered on
$(-0.11 \pm0.09^{\circ},-0.13 \pm0.08 ^{\circ})$.  
This is consistent with the Soudan~2~\cite{Cobb:1999mi} shadow, which was
offset by $0.1^{\circ}$E-W and $0.1^{\circ}$N-S.
The error and maximal value was found by drawing a \unit[1]{$\sigma$}
contour around the bin location of $\Lambda_{max}$.
The value of $I_s$ at this location was 0.13, in which 235.2 events were
removed by the moon, which is consistent with the expectation of 297.8
events removed.  
In the high statistics limit, the distribution of $\Lambda$ is the
same as for a $\chi^2_{\nu}$ distribution~\cite{Cash:1979vz}, where $\nu$
is the number of free parameters.  In this case there is only
one free parameter, $I_s$, which means $\Lambda$~=~30.9 has a 
significance of 5.6~$\sigma$.

A second method was used as a cross-check to find the significance of the observed deficit.
Moonless grids were created using the same background
method described in Sec.~\ref{sub:bkgnd}.  Searching through many moonless grids 
gave the probability of randomly finding a
moon-like deficit for a particular value of $\Lambda$.  Each $4^{\circ}
\times 4^{\circ}$ grid allowed 160,000 searches, for a total of 161.4 million
searches.  There were no searches that had $\Lambda > 23.0$ and the
distribution of $\Lambda$ followed the expected $\chi^2_1$
distribution.

\subsection{One Dimensional Shadow}\label{sub:1dms}
As a check on the resolution of the detector, a one dimensional moon
shadow search was performed.
The differential density of cosmic rays obscured by the moon, $\Delta
N_{\mu}/\Delta 
\Omega$,  can be written as
a two dimensional Gaussian convolution in polar coordinates (r,$\phi$):
\begin{equation}
 \frac{\Delta N_{\mu}}{\Delta \Omega} = \lambda \left( 1 - \frac{1}{2
 \pi\sigma^2} \int_0^{R_m} \! r' \, dr' \int_0^{2\pi} \! \, d\phi
~e^{-(r^2+r'^2-2r r'\cos\phi)/2\sigma^2 }\right)
\end{equation}
 where $\lambda$ is the average differential muon flux, $\sigma$ accounts
 for smearing from detector resolution, multiple Coulomb scattering and
 geomagnetic deflection, and $R_m = $0.26$^{\circ}$, the radius of
 the moon or sun.   
Performing the integration over $\phi$ gives:
\begin{equation}
 \frac{\Delta N_{\mu}}{\Delta \Omega} = \lambda \left( 1 - \frac{1}{
 \sigma^2} \int_0^{R_m} \! r' \, dr' 
\mathrm{I}_0\left(\frac{r r'}{\sigma^2} \right) e^{-\frac{r^2}{2\sigma^2} }\right),
\end{equation}
where $\mathrm{I}_0$ is a modified Bessel function of the first kind.
The integral that remains has no closed-form solution; however
$\mathrm{I}_0$ has a rapidly decreasing Taylor expansion:
\begin{equation}
 \mathrm{I}_0(x) = \sum^{\infty}_{k=0} \frac{2^{2k}}{4^k(k!)^2} = 1+ \frac{x^2}{4}+
 \frac{x^4}{64}+\frac{x^6}{2304}+ ...
\end{equation}
Substituting the expansion and evaluating the integral gives: 
 \begin{equation}\label{eq:fit}
 \frac{\Delta N_{\mu}}{\Delta \Omega} =
 \lambda \bigg[ 1-\frac{R_m^2}{2\sigma^2} e^{-\theta^2/2\sigma^2} \bigg(
 1+ \frac{(\theta^2-2\sigma^2)R^2_M}{8\sigma^4}
 + \frac{(\theta^4-8\theta^2\sigma^2+8\sigma^4)R^4_M}{192\sigma^8}\bigg) \bigg],
 \end{equation}
with $\theta = r$ since the observable in this case is an angular separation.  
This formula automatically produces a deficit of $\pi R^2_m \lambda$ events due to shadowing.
The significance of
the deficit can be found by fitting to Eq.~\ref{eq:fit} and finding the
difference between this $\chi^2$ value and the $\chi^2$ value obtained
by a linear fit to the same data.   

\subsection{Search in One Dimension} 
The reconstructed muon angular separation from the moon or sun, $\Delta
\theta$, was binned in $S_\mathrm{bin} = $0.10$^{\circ}$ increments.  
Since radial distance from the center is measured over
a two dimensional projection,
the solid angle of bin (i) increases when moving out from the center
as $\Delta \Omega_i = (2i-1) \cdot S^2_\mathrm{bin}\pi$.  Weighting the number of events in each
bin by the reciprocal of the area resulted in the distribution
$N_i/\Delta \Omega_i$, the differential muon density. 

\begin{figure}[h!]
\begin{center}
\includegraphics[width=0.8\textwidth]{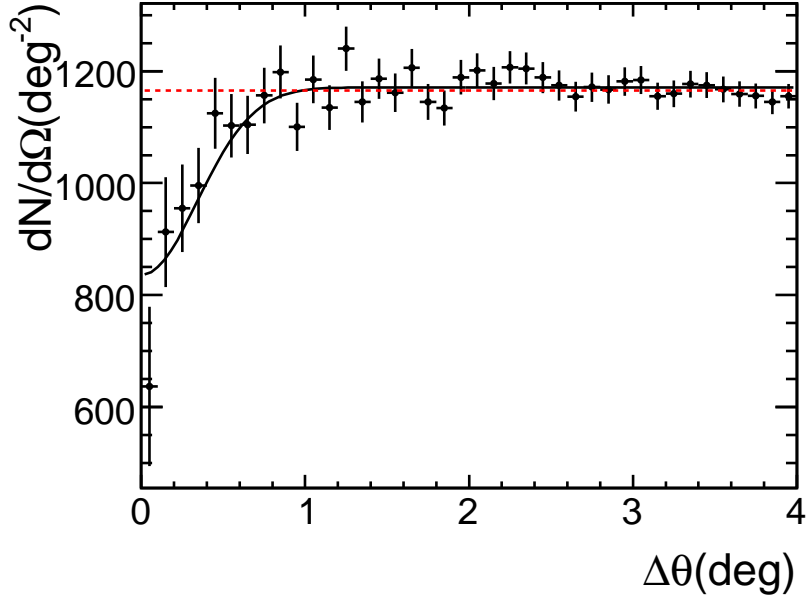}
\vspace{-10pt}
\caption[One Dimensional Moonshadow]{\label{fig:msData} The differential muon flux with respect to
displacement from the moon's location, 
binned in $\Delta \theta = 0.1^{\circ}$.  The
dashed curve is the result of a linear
(no moon effect) fit, which gives 
 $\chi^2_{L}/ndf = 57.8/39$.  The solid curve is the best fit from
 Eq.~\ref{eq:fit}.  The Gaussian (moon-induced deficit) fit gives
 $\chi^2_{G}/ndf = 25.3/38$,  with parameters $\lambda = 1171.1\pm4.8$
 and $\sigma = 0.34\pm0.04^{\circ}$.} 
\end{center}
\end{figure}

Since the location of the moon was found to be offset from
$(0^{\circ},0^{\circ})$, the arrival direction of each 
muon was adjusted by $(0.11^{\circ},0.13^{\circ})$ before performing the
one dimensional moon shadow search.
The position of the moon on the sky and the separation of each muon from the moon
($\Delta \theta$), was found using the method described in Sec.~\ref{sec:event_sel}.
The $\Delta \theta$ distribution is shown in Fig.~\ref{fig:msData} with
statistical error bars.  
There is a significant deficit in the location of the moon as $\Delta
\theta~\rightarrow~0$ attributed to 
the moon's blocking of the primary cosmic rays. A linear 
(no moon deficit) 
fit gives a $\chi^2_{L}/ndf=57.8/39$.  A fit to
Eq.~\ref{eq:fit} ( moon induced deficit) gives a
$\chi^2_{G}/ndf=25.3/38$. The improvement in 
$\chi^2$ of 32.6
corresponds to a 
significance of 5.6~$\sigma$.  
The resolution
found by this method is $\sigma = 0.34\pm0.04^{\circ}$, 
which is consistent with a Gaussian fit to the multiple muon data shown in Fig.~\ref{fig:resolution}.
When the same analysis is
performed without shifting the muons by $(0.11^{\circ},0.13^{\circ})$,
the significance of the moon shadow is reduced by  
$2.7~\sigma$, and
the shadow is more smeared out, with $\sigma = 0.39 \pm
0.06^{\circ}$.  

\section{Sun Shadow}\label{sec:sunshadow}
As viewed from earth, the sun also obscures a
$0.52^{\circ}$  diameter disk, approximately the same size as the moon. 
The one dimensional shadowing procedure from Sec.~\ref{sec:shadowing}
and the two dimensional log-likelihood analysis described in
Sec.~\ref{sub:2dms} were performed for cosmic ray muons coming from
the direction of the sun.
Historically, this has been a more difficult~\cite{Ambrosio:2003mz}
measurement to make because of the variability of the sun's magnetic
field described in Sec.~\ref{sec:imfeffects}.

\begin{figure}[h!]
\begin{center}
\ifbw
\includegraphics[width=0.8\textwidth]{2DSunAll-BW}
\else
\includegraphics[width=0.8\textwidth]{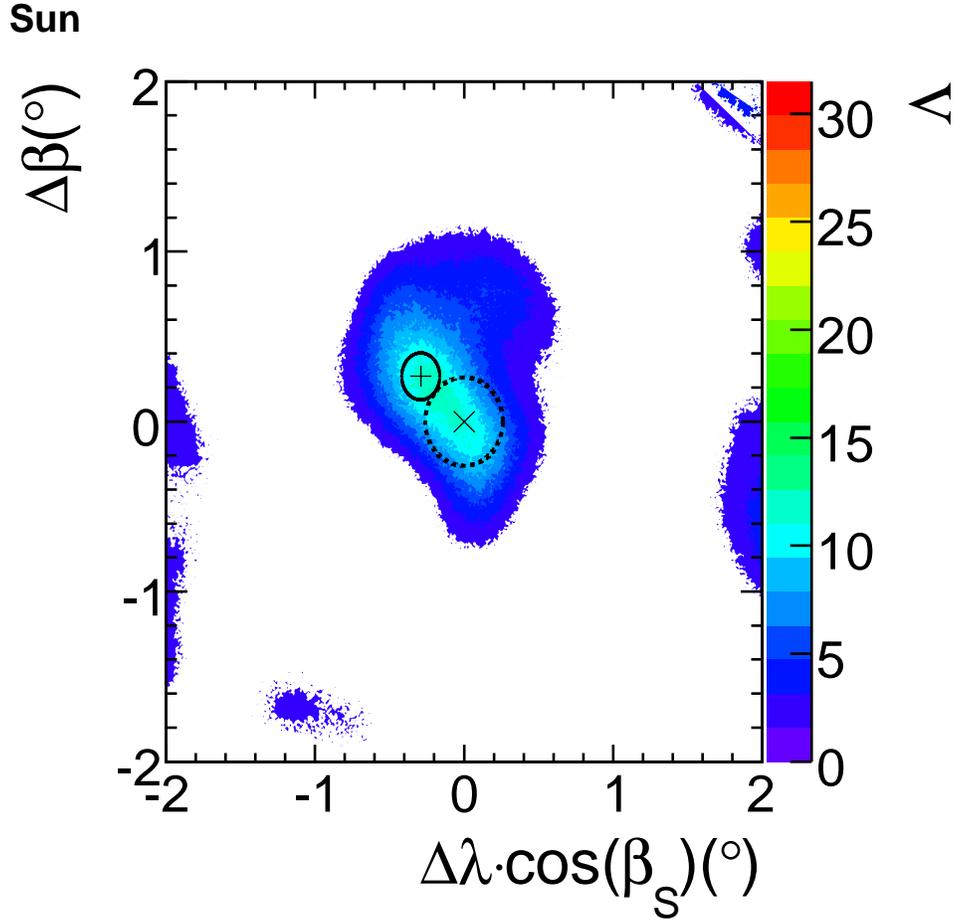}
\fi
\caption[Two Dimensional Sunshadow]{\label{fig:2DSunShadow} The two dimensional sun induced muon
deficit in $\unit[0.01]{deg^2}$ bins, in ecliptic coordinates.  The $\times$
marks the expected location of the sun, and the dashed circle is the apparent
size of the sun as viewed from earth.  The cross marks the observed
location of the sun, and the solid ellipse denotes the uncertainty.  The greatest deficit is $\Lambda_{max} = 14.6$, 
a \unit[3.8]{$\sigma$} significance,
centered on $(-0.29^{\circ},
0.27^{\circ}$).  The maximum of the color scale is set to $\Lambda = 31$
to allow for easy comparison with the moon shadow.}
\end{center} 
\end{figure}
The two dimensional sun shadow can be seen in Fig.~\ref{fig:2DSunShadow}.
The sun $\Lambda_{max} = 14.6$  occurs at $(-0.29\pm0.13^{\circ},
0.27 \pm 0.14^{\circ}$) and has a~3.8$\sigma$ significance.  
The value of $I_s$ at this location was 0.08,
in which 144.6 events were 
removed by the sun, which is about half of the expected of 279.8
events removed.   
The sun shadow has a $\Lambda_{max}$ value that
is less than half the $\Lambda_{max}$ value of the moon, and the shadow
appears elongated along a line through the origin, rotated about halfway
between the north and 
to the west.  These features could be attributed to 
a smearing effect that is
unobserved in the moon shadow, originating from the longer exposure to the
IMF along the much greater path between the sun and the earth. 

\begin{figure}[h!]
\begin{center}
\includegraphics[width=0.8\textwidth]{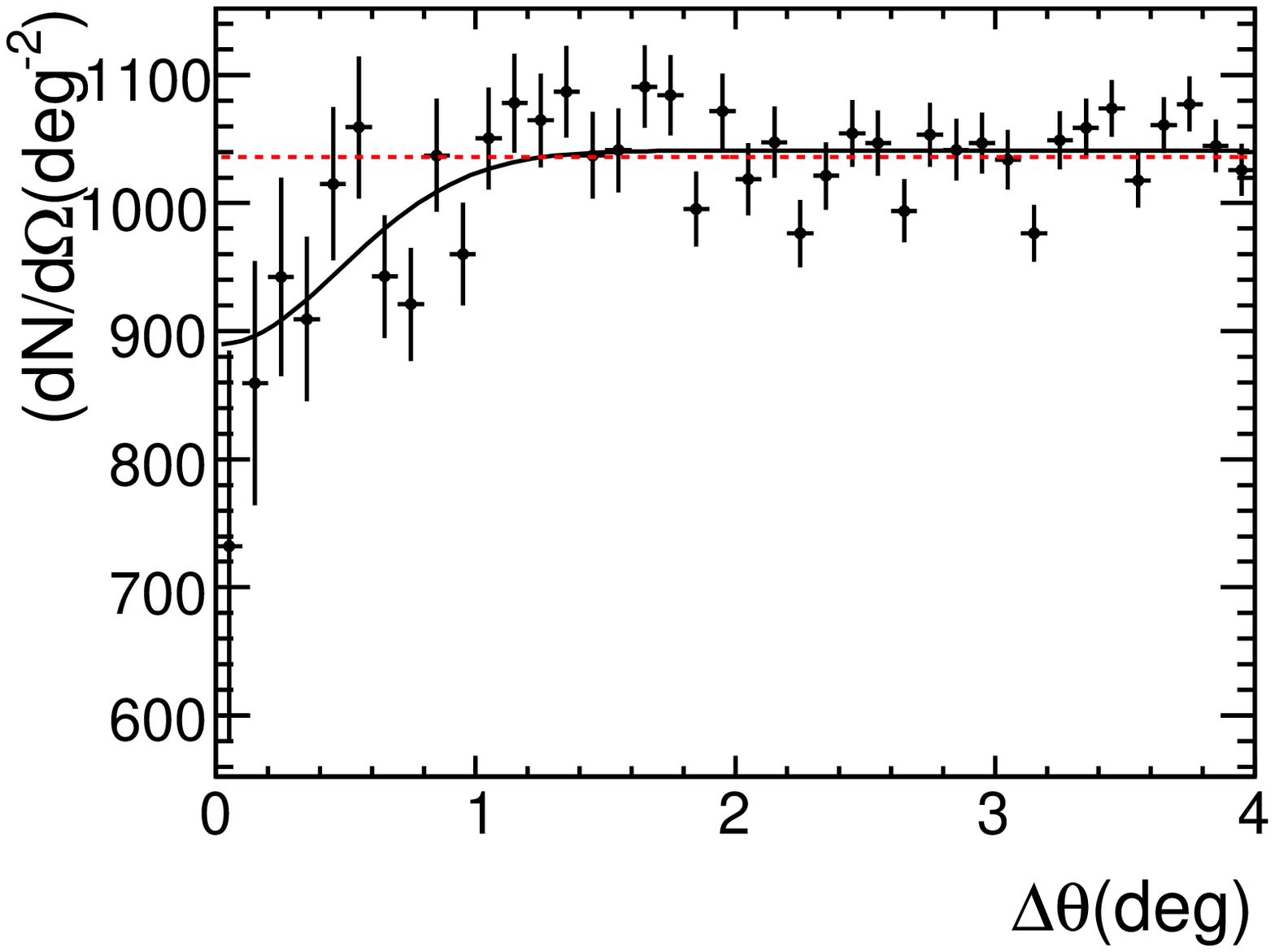}
\caption[One Dimensional Sunshadow]{\label{fig:sunData} The differential muon flux with respect to
displacement from the sun's location, 
binned in $0.10^{\circ}$.  
A linear (no sun effect) fit gives 
$\chi^2_{L}/ndf = 67.9/39$.  The solid curve is the best fit from
Eq.~\ref{eq:fit}.  The Gaussian (sun-induced deficit) fit gives
$\chi^2_{G}/ndf = 52.8/38$,  with parameters $\lambda = 1040.9\pm4.6$
and $\sigma = 0.48\pm0.073^{\circ}$.} 
\end{center}
\end{figure}

The one dimensional sun shadow can be seen in Fig.~\ref{fig:sunData}.
Note that the muons were shifted by $(0.29^{\circ},-0.27^{\circ})$ to
account for the offset from the expected location of the sun given by
the two dimensional sun shadow.
There is a significant deficit in the location of the sun attributed
to the sun's blocking of the primary cosmic rays. The improvement in
$\chi^2$ of 15.1 ($\chi^2_{L}/ndf = 67.9/39, \chi^2_{G}/ndf = 52.8/38$)
corresponds to a 
3.9~$\sigma$ significance .  The Gaussian
fit parameter $\sigma = 0.48\pm0.07$ for the sun shadow is somewhat
larger than the value found by the fit to the moon shadow,
$\sigma = 0.34 \pm0.04$. 

\section{Interplanetary Magnetic Field Effects}\label{sec:imfeffects}
\subsection{Moon Shadow}
\begin{figure}[!h]
\begin{center}
\begin{minipage}[l]{0.49\linewidth}
\ifbw
\includegraphics[width=0.99\textwidth]{2DMSAll-night-BW}
\else
\includegraphics[width=0.99\textwidth]{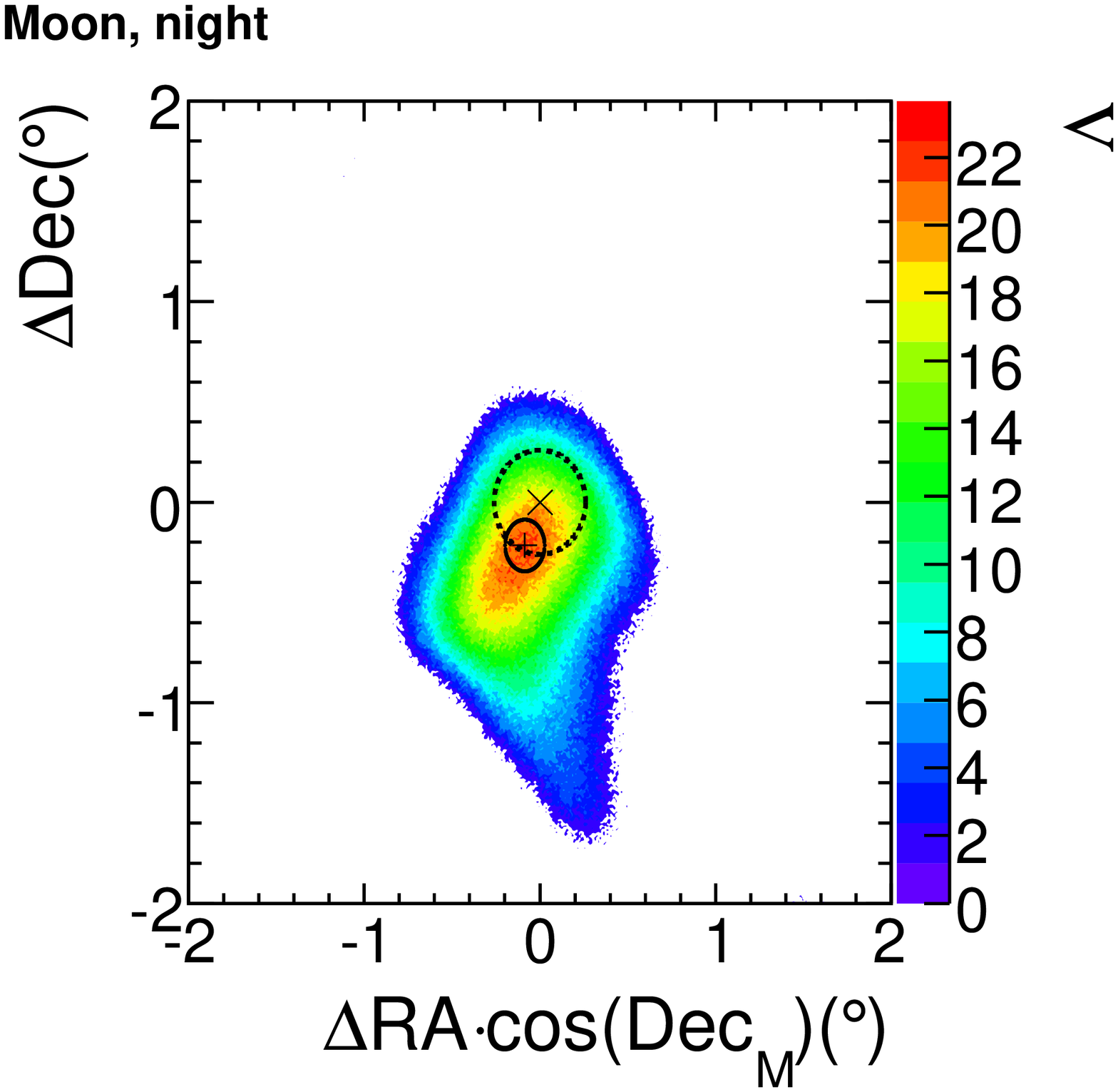}
\fi
\end{minipage}
\begin{minipage}[r]{0.49\linewidth}
\ifbw
\includegraphics[width=0.99\textwidth]{2DMSAll-day-BW}
\else 
\includegraphics[width=0.99\textwidth]{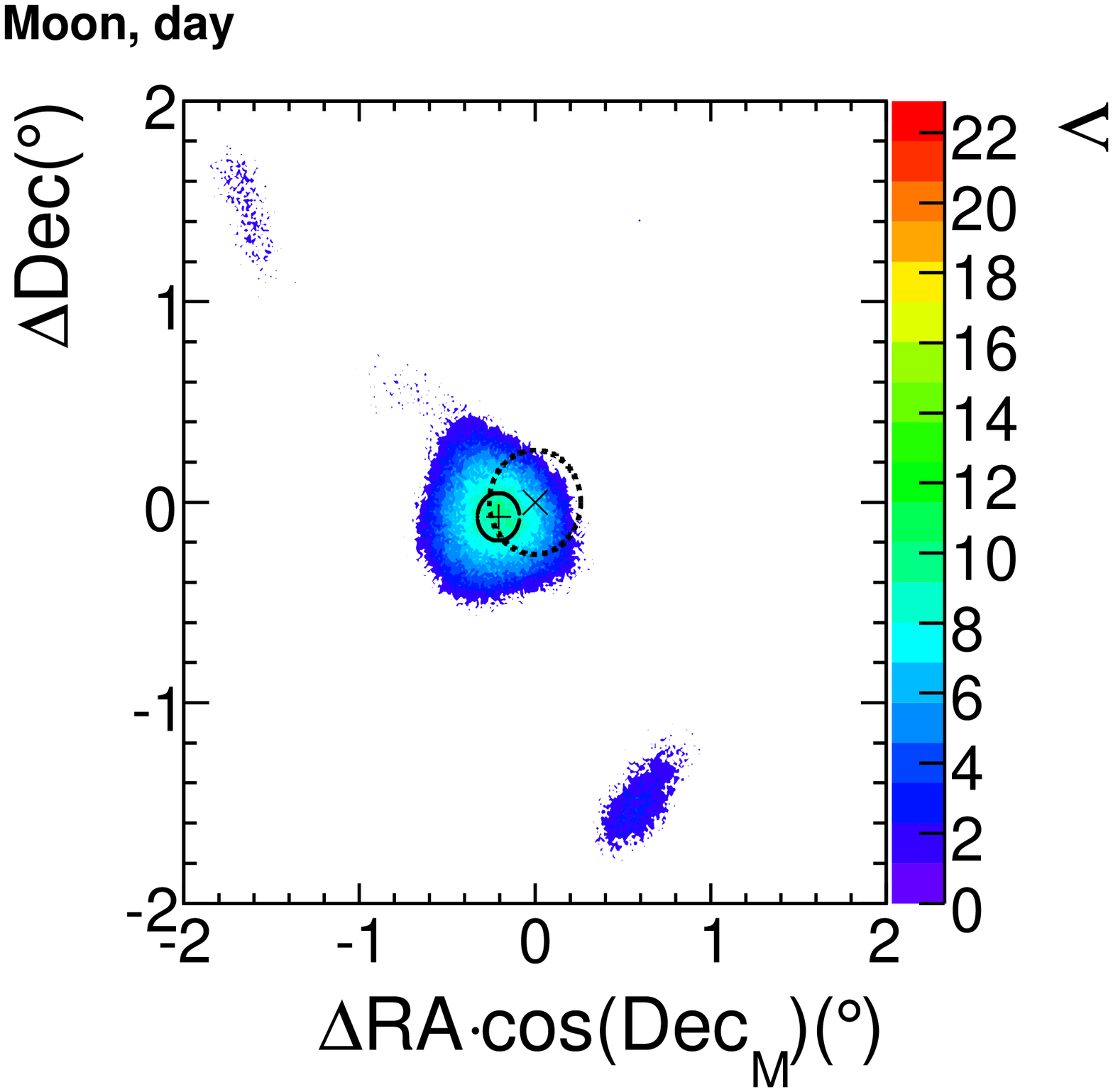}
\fi
\end{minipage}
\end{center}
\caption[Two Dimensional Moonshadow-Night]{\label{fig:MSAll-Day}  The distribution of $\Lambda$ values in
celestial coordinates when the moon was visible during the night~(Left),
and when the moon was visible during the day~(Right).   The $\times$
marks the expected location of the moon, and the dashed circle is the apparent
size of the moon as viewed from earth. The cross marks the observed
location of the moon, and the solid ellipse denotes the uncertainty. The
deficit has $\Lambda_{max}$=23.7 (\unit[4.9]{$\sigma$}) at night, and the center
is at $(-0.09\pm0.11^{\circ},-0.22\pm0.13^{\circ})$. The
deficit has $\Lambda_{max}$=11.2 (\unit[3.3]{$\sigma$}) in the day, and the center is
at $(-0.21\pm0.12^{\circ},-0.07\pm0.12^{\circ})$. }  
\end{figure}

The IMF  could have some effect on
the moon shadow by the effects described in Sec.~\ref{sec:intro}.  This could be observed by dividing the data into
separate day/night samples as cosmic rays experience the maximal IMF difference when approaching the earth from the day and night sides.
The effect of the IMF on the geomagnetic field is to produce a reduced geomagnetic field at night ~\cite{Tsyganenko:1995,Tsyganenko:1996,Zhou:1997}.  This effect was
observed by MACRO~\cite{Ambrosio:2003mz}.  Daytime is defined here as when the sun's
zenith angle was less than zero.  The day-time sample contained 8,270 muons within a
$2^{\circ}$ half-angle cone around the 
moon, while the night-time sample contained 9,213 events.  The reason for this difference is that the detector is only down
for maintenance during the day, 
coupled with the fact that more of the data were collected in the
fall and winter than spring and summer (see Sec.~\ref{sec:event_sel}).  Integrating over five and a half years accounts for the 10\%
increase in muons collected near the moon at night.  The moon shadow
observed at night can
be seen in Fig.~\ref{fig:MSAll-Day}~(Left), while the moon shadow during the
day can
be seen in Fig.~\ref{fig:MSAll-Day}~(Right). 
 The center of the deficit is at
 $(-0.09\pm0.11^{\circ},-0.22\pm0.13^{\circ})$  with
 $\Lambda^{night}_{max} = $  
23.7 (\unit[4.9]{$\sigma$}) for the data taken at
 night.  For the data taken during the day, the center of the deficit is
at $(-0.21\pm0.12^{\circ},-0.07\pm0.12^{\circ})$, with
$\Lambda^{day}_{max} = $ 11.2 (\unit[3.3]{$\sigma$}).  
$\Lambda^{night}_{max}$ 
is consistent with the expected 
location of the moon in $\Delta RA$, while the shadow during the day is
shifted by \unit[1.7]{$\sigma$}.  The
shift in $\Delta RA$ at night is consistent with zero.  The shift in
$\Delta Dec$ at night, however, is further from the expected location of
the moon by \unit[1.7]{$\sigma$}. 
It is unlikely that the geomagnetic
field could cause a deflection of greater than $-0.02^{\circ}$ (
Sec.~\ref{sub:geomag}) and the geomagnetic field should not cause a
diurnal effect.  A possible explanation is that the IMF 
has a general defocussing effect on cosmic rays, and cosmic rays
detected at night experience less of this IMF effect because they arrive
at earth from the opposite direction to the sun.

\subsection{Sun Shadow}
\begin{figure}[!h]
\begin{center}
\begin{minipage}[l]{0.49\linewidth}
\includegraphics[width=0.99\textwidth]{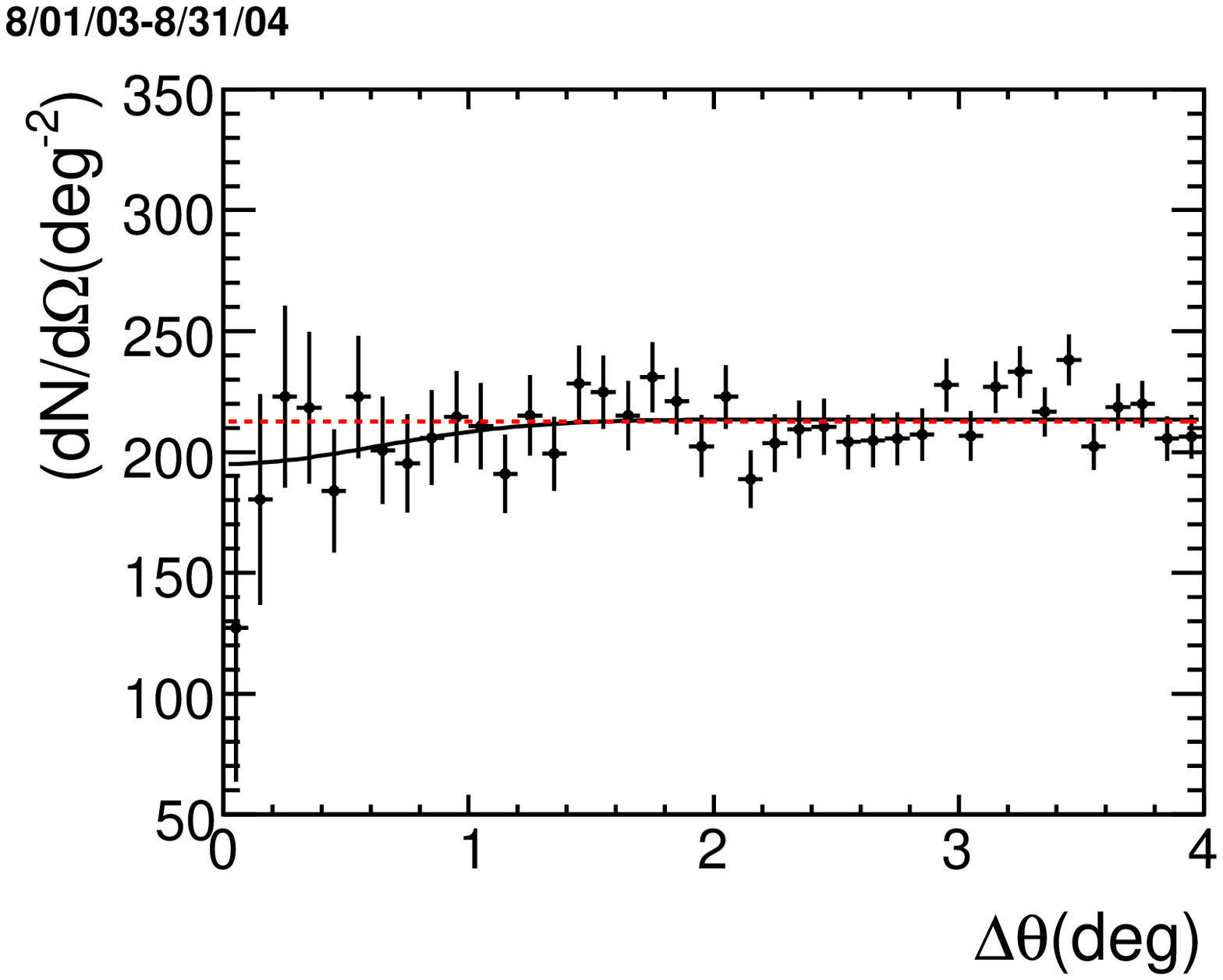}
\end{minipage}
\begin{minipage}[r]{0.49\linewidth}
\includegraphics[width=0.99\textwidth]{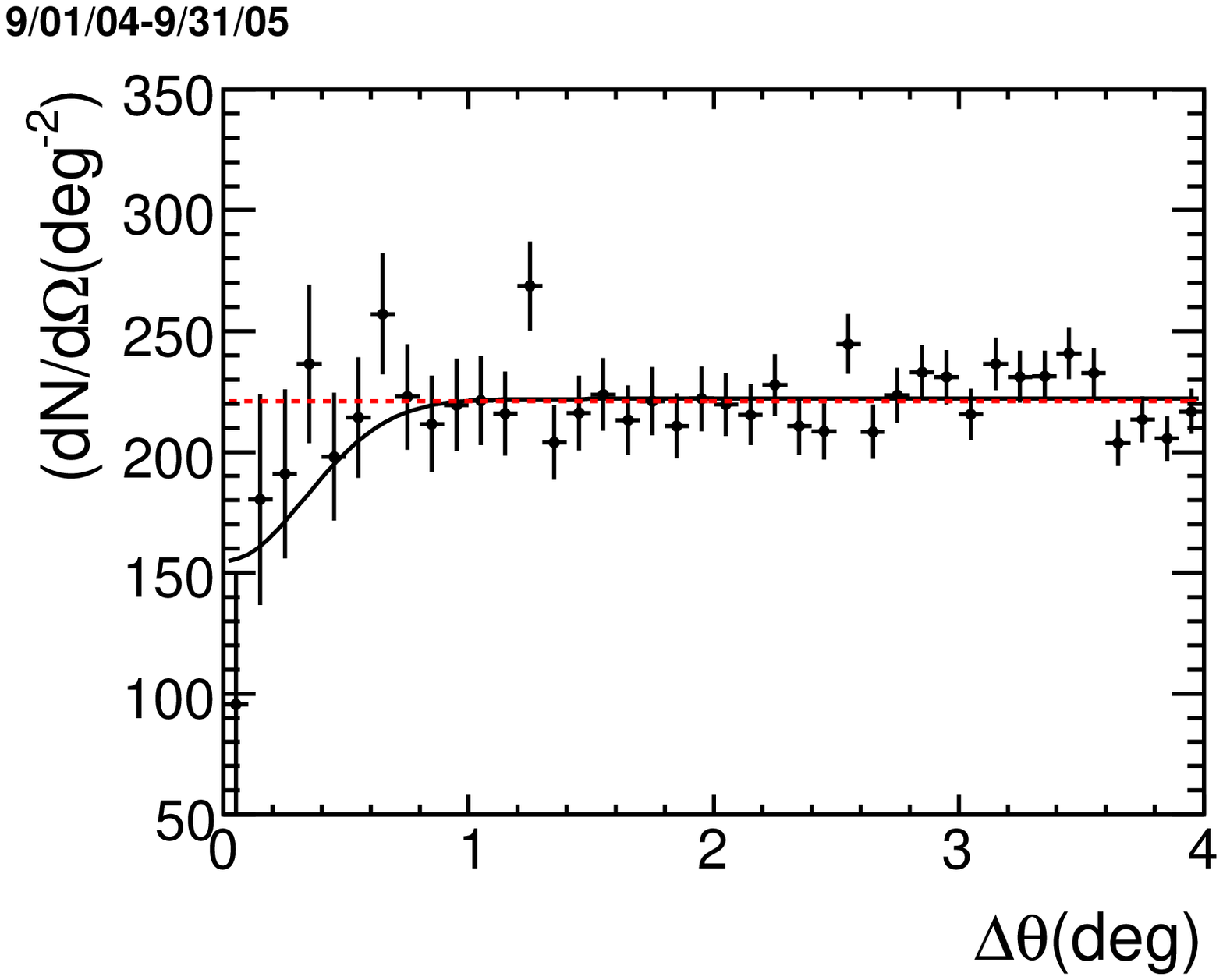}
\end{minipage}
\begin{minipage}[l]{0.49\linewidth}
\includegraphics[width=0.99\textwidth]{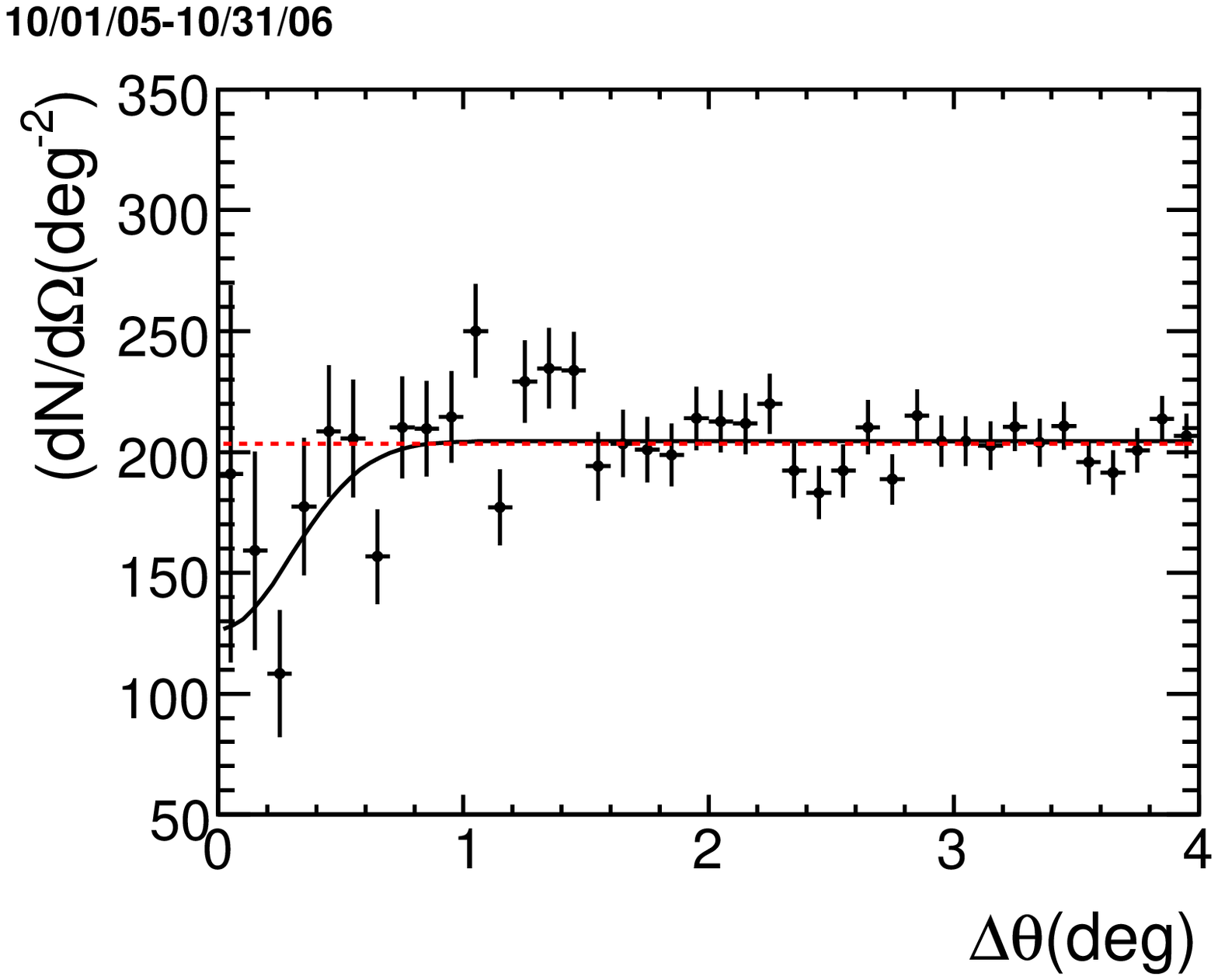}
\end{minipage}
\begin{minipage}[r]{0.49\linewidth}
\includegraphics[width=0.99\textwidth]{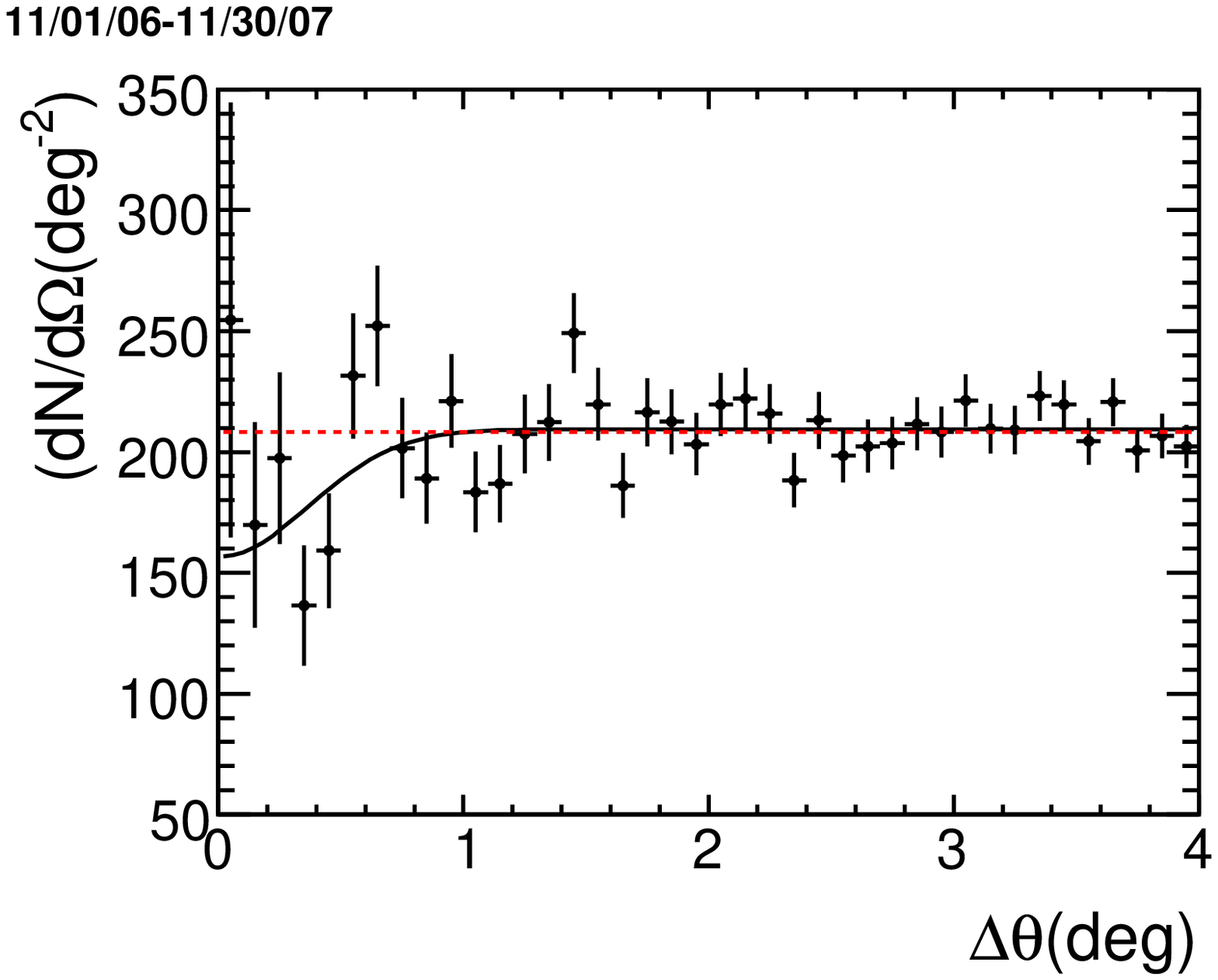}
\end{minipage}
\includegraphics[width=0.49\textwidth]{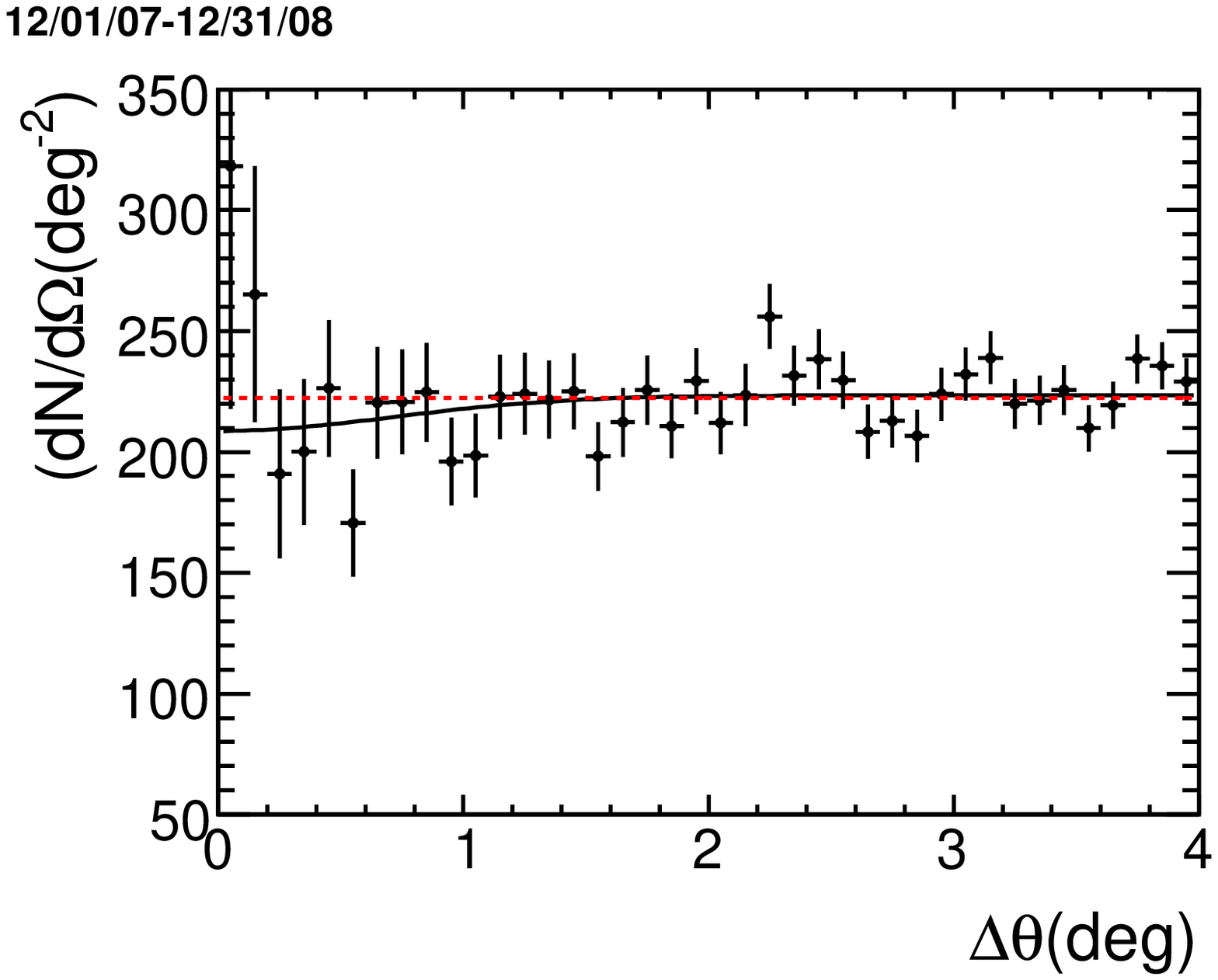}
\caption[Sunshadow Per Year]{\label{fig:SunShadowYear} The differential muon flux with respect to
  displacement from the sun's location as a function of year, 
  binned in $0.1{^o}$.  The dashed curve is a linear
  (no sun effect) fit, while the solid curve
  is the best fit from Eq.~\ref{eq:fit}.
}
\end{center}
\end{figure}

A greater IMF
means a less prominent sun shadow, so the significance of the observable
shadowing caused by the sun should decrease as the IMF increases.  Since
the IMF is caused by solar activity, the significance of observable
shadowing should increase as the number of sunspots, one measure of
solar activity, decreases.
The last maximum of the 11~year solar cycle occurred in 2001, and the
next minimum occurred in December,~2008~\cite{SolarCycle}.
 To search for
a correlation between solar activity and strength of sun shadowing
effects, the data were divided into five separate one dimensional grids
of roughly equal statistics, and fit with both a linear function and a
one-parameter Gaussian (Eq.~\ref{eq:fit}), holding the number of
removed muons constant.
The one dimensional search is used because
it requires fewer events to produce a statistically significant shadow
and the sun was shown to be close to the center of the grid (see
Fig.~\ref{fig:2DSunShadow}).  
There were not enough accumulated muons to perform five of the two
dimensional likelihood searches.  The periods were
Aug.~1,~2003~-~Sep.~30,~2004, Oct.~1,~2004~-~Oct.~31,~2005,
Nov,~1,~2005~-~Nov.~30,~2006, Dec.~1,~2006~-~Dec.~31,~2007 and Jan.~1,~2008~-~Dec.~31,~2008. These
graphs can be seen in Fig.~\ref{fig:SunShadowYear}.
\begin{table} [!h]
  \centering
  \caption[Yearly Sunshadow Significance] { 
    The size of shadowing
    observed in each year's sun
    distribution.
    Also included is the corresponding moon shadow size.  The
    significance is given in units of standard deviations (s.d.). 
    \label{tab:sigtable} }
  \begin{tabular}{||l|c|c|c|c|c|c||} \hline\hline
    & \multicolumn{3}{|c|}{Sun}  &  \multicolumn{3}{|c||}{Moon}\\\hline\hline 
    \textbf{Distribution}  &   $\sigma$ 
    & $\chi_{G}^2$ 
    &  \textbf{Significance }  
    & $\sigma$ 
    & $\chi_{G}^2$ &
    \textbf{Significance } 
    \\ \hline \hline
    
    Aug. 1, 2003 - Sep. 30, 2004 &0.61 $\pm$ 0.28 & 34 &\unit[1.6]{s.d.} & 0.34 $\pm$ 0.11 & 27.4 &  \unit[ 1.9 ]{s.d.}\\ \hline
    Oct. 1, 2004 - Oct. 31, 2005 &0.31 $\pm$ 0.12 & 42.3 &  \unit[1.6]{s.d.} & 0.33 $\pm$ 0.08 & 24 &  \unit[ 4.0 ]{s.d.}\\ \hline
    Nov. 1, 2005 - Nov. 30, 2006 &0.27 $\pm$ 0.07 & 45.5 &  \unit[3.2]{s.d.} & 0.19 $\pm$ 0.04 & 38.6 &  \unit[ 3.9 ]{s.d.}\\ \hline
    Dec. 1, 2006 - Dec. 31, 2007 &0.34 $\pm$ 0.12 & 42.3 &  \unit[2.5]{s.d.} & 0.33 $\pm$ 0.11 & 32.2 &  \unit[ 3.6 ]{s.d.}\\ \hline
    Jan. 1, 2008 - Dec. 31, 2008 &0.70 $\pm$ 0.29 & 37.2 &  \unit[2.1]{s.d.} & 0.26 $\pm$ 0.06 & 42.6 &  \unit[ 3.4 ]{s.d.}\\ \hline
  \end{tabular}
\end{table}

The results of these fits are summarized in Table~\ref{tab:sigtable}.
There is a decrease in the size ($\sigma$) and corresponding error of the sun shadow
as the sun approaches
solar minimum.  This decline could be correlated to the decrease in solar
activity.  Surprisingly, the period nearest to solar minimum, January 1,
2008-December 31, 2008, shows an increased size. 
For comparison, the yearly moon shadows for the same periods were
calculated and summarized in Table~\ref{tab:sigtable}.  The moon shadow
has more constant size over the five periods than does the sun.

A more useful correlation between the sun shadow and the IMF is to
perform two dimensional searches on data sets subdivided according to 
the vector components of the IMF, but there are problems with this approach.
First of all, the number of data subsets is limited to two, because the
2D log-likelihood analysis is statistically limited.  Second, the IMF is
known to have a complicated and time-varying structure.  The 
ACE, Wind and IMP 8 spacecraft, which monitor the IMF
while orbiting about the L1 Lagrange point (\unit[225]{earth radii}  
 in front of earth), report data averaged over one hour~\cite{omniweb}.
These data report the IMF in Geocentric Solar Ecliptic (GSE)
coordinates, where the earth-sun line defines the x-axis, and the
ecliptic north pole is the z-axis. 
It is difficult  to adequately account for the variations in the field
with these data, as was done in 
Sec.~\ref{sub:geomag} for the geomagnetic field.  The greatest
difficulty is the fact that the
IMF data is only available every hour for one location in space, whilst
it is well known that the IMF changes somewhat rapidly in time and has a
sectorized structure~\cite{Wilcox:1965}.  The direction of the IMF is
either ``toward'' or ``away'' from the earth, and the two field
directions are separated by a thin layer known as the neutral current
sheet, which passes the L1 point very briefly as the $x$ component of
IMF reverses polarity.

These IMF
data were used to separate the muon data and search for IMF induced
effects.  The data were first separated into two sets by the magnitude
in the expected deflection plane, which is perpendicular to the
earth-sun line.  The ``large magnitude'' set was where
$|B_{\perp}|>\unit[4.2]{nT}$; less than \unit[4.2]{nT} was
the ``small magnitude'' data set.  No significant translation of the
shadow nor focusing effects were observed.  The shadows had magnitudes
equal to roughly half of the total sun shadow, consistent with what was
expected from dividing the data set in half arbitrarily.  

An interesting
effect was observed when the data were divided into sets where the
IMF component along the earth-sun line ($B_{x}$) was greater and less
than zero.  
For the $B_x>0$ case, the most significant
deficit was centered on ($-0.07\pm0.14^{\circ},
0.10\pm0.12^{\circ}$), which is consistent with the expected location of
the sun.  This shadow shows none of the elongation seen in the total sun
shadow (Fig.~\ref{fig:2DSunShadow}).  For the $B_x<0$ case, the most significant
deficit was centered on ($-0.41\pm0.17^{\circ},
0.47\pm0.20^{\circ}$).  In fact, this shadow has almost no significance
near the expected location of the sun, and has two deficits of almost
equal significance an equal distance from the center.  This offset is
\unit[2.4]{$\sigma$} in RA and \unit[2.3]{$\sigma$} in Dec from the location of the
sun.
\begin{figure}[!h]
\begin{center}
\begin{minipage}[l]{0.49\linewidth}
\ifbw
\includegraphics[width=0.99\textwidth]{2DSunAll-imfxpos-BW}
\else
\includegraphics[width=0.99\textwidth]{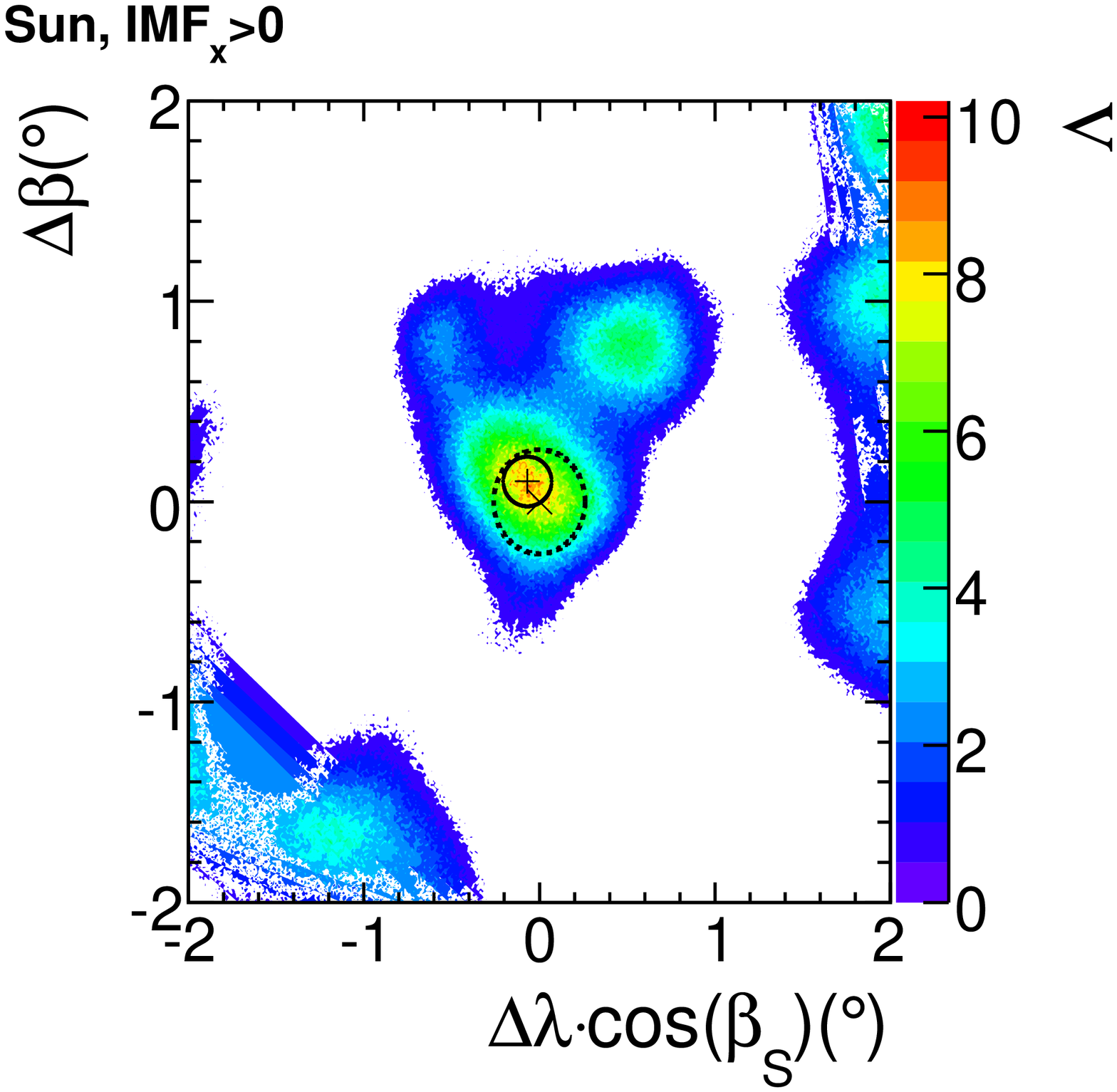}
\fi
\end{minipage}
\begin{minipage}[r]{0.49\linewidth}
\ifbw
\includegraphics[width=0.99\textwidth]{2DSunAll-imfxneg-BW}
\else
\includegraphics[width=0.99\textwidth]{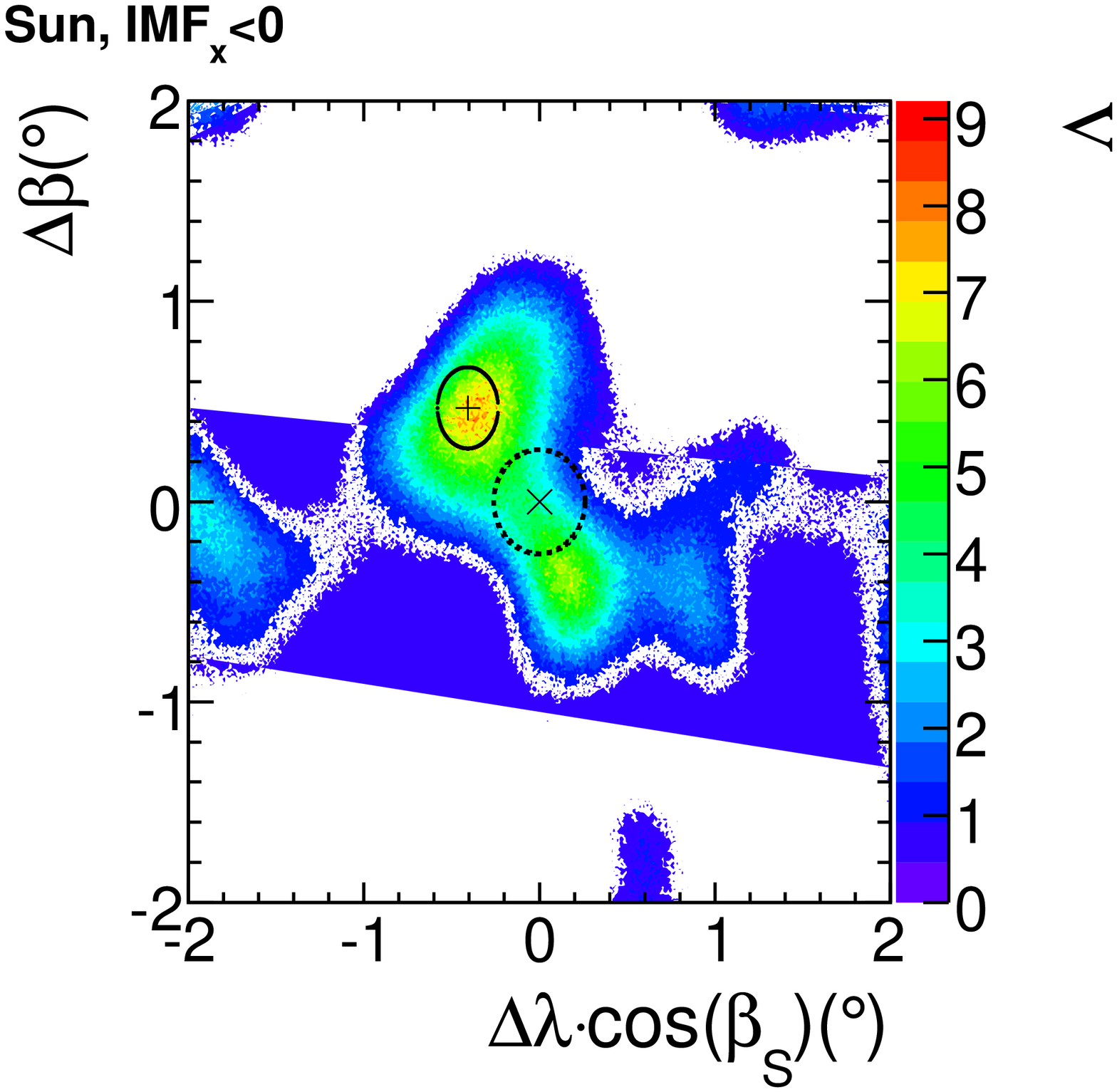}
\fi
\end{minipage}
\caption[Sun Shadow for B0 and B0]{\label{fig:sunimf}  
The sun shadow for $B_x > 0$ (Left) and $B_x < 0$ (Right).  The $\times$
marks the expected location of the sun, and the dashed circle is the apparent
size of the sun as viewed from earth. The cross marks the observed
location of the sun, and the solid ellipse denotes the uncertainty.
  The most significant 
deficit was located at ($-0.07\pm0.14^{\circ}, 0.10\pm0.12^{\circ}$) 
and had $\Lambda_{max} = 10.2$ (\unit[3.2]{$\sigma$}) for $B_x > 0$.  The 
most significant deficit was located at ($-0.41\pm0.17^{\circ}, 
0.47\pm0.20^{\circ}$) and had $\Lambda_{max} = 9.2$
(\unit[3.0]{$\sigma$}) for $B_x < 0$.}
\end{center}
\end{figure}
The sectorized structure of the IMF causes a cosmic ray traveling along the
path from the sun to the earth to experience magnetic deflections in a
multitude of directions.  The rotating nature of the IMF adds an
additional element to the magnetic deflection.  It may be that the
``inward'' and ``outward'' separation of the cosmic ray data set along the
earth-sun line is a
more consistent predictor of cosmic ray deflection than the sparse
measurements of the IMF in the y and z directions.

\section{Conclusions}
Using  52.3 million muons accumulated over 1857.91 live-days, the
MINOS far detector has observed the cosmic ray shadow of both the moon and sun with
high significance.  
The two dimensional moon shadow was seen with a significance of
5.6~$\sigma$, centered on
$(-0.11\pm0.09^{\circ},-0.13\pm0.08^{\circ})$, which suggests that the
absolute pointing of the detector on the sky is known to $0.17\pm
0.12^{\circ}$.  The significance of the one dimensional moon shadow
increased after shifting the muon arrival directions by
$(-0.11^{\circ},-0.13^{\circ})$ to be consistent with the observed
location of the moon.  The cosmic ray shadow of the sun over the
same time period was measured 
in two dimensions with a
significance 3.8~$\sigma$, centered on
 $(-0.29 \pm0.13^{\circ}, 
0.27\pm0.14^{\circ}$).  

Searches were made for possible dependencies of the moon and sun shadow on the
IMF but no three sigma effects were observed.  At lesser statistical
significance, effects were observed in the difference of the moon shadow during
the day and night and in the variation of the sun shadow with time.
The shadow of the sun showed some indication (though less than three sigma) of variation with the x
component of the IMF, which could be interpreted as focusing (for $B_x>0$)
and defocussing (for $B_x<0$).  These latter observations of short term correlations with the IMF were made at the
limit of the experiment's statistical power, and it will be interesting
for other detectors to confirm these signals.

\section{Acknowledgments}
We thank the Fermilab staff and the technical staffs of the participating institutions for their vital contributions. This work was supported by the U.S. Department of Energy, the U.S. National Science Foundation, the U.K. Science and Technologies
Facilities Council, the State and University of Minnesota, the Office of
Special Accounts for Research Grants of the University of Athens,
Greece, FAPESP (Funda\c c\~ao de Amparo \`a Pesquisa do Estado
de S\~ao Paulo), CNPq (Conselho Nacional de Desenvolvimento
Cient\'ifico e Tecnol\'ogico) in Brazil.  We gratefully acknowledge the Minnesota Department of Natural
Resources for their assistance and for allowing us access to the
facilities of the Soudan Underground Mine State Park and the crew
of the Soudan Underground Physics laboratory for their tireless work in
building and operating the MINOS detector. 
\bibliography{ShadowsAstroPart}

\begin{thebibliography}{10}
\expandafter\ifx\csname url\endcsname\relax
  \def\url#1{\texttt{#1}}\fi
\expandafter\ifx\csname urlprefix\endcsname\relax\def\urlprefix{URL }\fi

\bibitem{MinosNIM}
D.~G. Michael, et~al., The magnetized steel and scintillator calorimeters of
  the {MINOS} experiment, Nucl. Instrum. Methods 596 (2008) 190--228.

\bibitem{Adamson:2007gu}
P.~Adamson, et~al., {A Study of Muon Neutrino Disappearance Using the Fermilab
  Main Injector Neutrino Beam}, Phys. Rev. D77 (2008) 072002.

\bibitem{Rebel:2004mm}
B.~J. Rebel, {Neutrino induced muons in the {MINOS} far detector},{P}hD
  Dissertation, Indiana University {F}ERMILAB-THESIS-2004-33.

\bibitem{Adamson:2009zf}
P.~Adamson, et~al., {Observation of muon intensity variations by season with
  the {MINOS} far detector}, {Submitted to Phys. Rev. D, hep-ex/09094012}.

\bibitem{Ambrosio:2002wr}
M.~Ambrosio, et~al., Search for cosmic ray sources using muons detected by the
  {MACRO} experiment, Astropart. Phys. 18 (2003) 615--627.

\bibitem{Clark:1957}
G.~W. {Clark}, {Arrival Directions of Cosmic-Ray Air Showers from the Northern
  Sky}, Physical Review 108 (1957) 450--457.

\bibitem{Alexandreas:1990wj}
D.~E. Alexandreas, et~al., Observation of shadowing of ultrahigh-energy cosmic
  rays by the moon and the sun, Phys. Rev. D43 (1991) 1735--1738.

\bibitem{Borione:1993xq}
A.~Borione, et~al., Observation of the shadows of the {M}oon and {S}un using
  100-{TeV} cosmic rays, Phys. Rev. D49 (1994) 1171--1177.

\bibitem{Amenomori:1993iv}
M.~Amenomori, et~al., Cosmic ray deficit from the directions of the {M}oon and
  the {S}un detected with the {T}ibet air shower array, Phys. Rev. D47 (1993)
  2675--2681.

\bibitem{Atkins:1999gb}
R.~W. Atkins, et~al., {Milagrito, a {TeV} air-shower array}, Nucl. Instrum.
  Meth. A449 (2000) 478--499.

\bibitem{Grapes}
A.~{Oshima}, S.~R. {Dugad}, U.~D. {Goswami}, S.~K. {Gupta}, Y.~{Hayashi},
  N.~{Ito}, A.~{Iyer}, P.~{Jagadeesan}, A.~{Jain}, S.~{Kawakami},
  M.~{Minamino}, P.~K. {Mohanty}, S.~D. {Morris}, P.~K. {Nayak}, T.~{Nonaka},
  S.~{Ogio}, B.~S. {Rao}, K.~C. {Ravindran}, H.~{Tanaka}, S.~C. {Tonwar},
  {GRAPES-3 Collaboration}, {The angular resolution of the GRAPES-3 array from
  the shadows of the Moon and the Sun}, Astroparticle Physics 33 (2010)
  97--107.

\bibitem{Hegra}
M.~{Merck}, A.~{Karle}, S.~{Martinez}, F.~{Arqueros}, K.~H. {Becker},
  M.~{Bott-Bodenhausen}, R.~{Eckmann}, E.~{Faleiro}, J.~{Fernandez},
  P.~{Fernandez}, V.~{Fonseca}, V.~{Haustein}, G.~{Heinzelmann}, I.~{Holl},
  F.~{Just}, F.~{Krennrich}, M.~{K{\"u}hn}, E.~{Lorenz}, H.~{Meyer},
  N.~{M{\"u}ller}, R.~{Plaga}, J.~{Prahl}, M.~{Probst}, M.~{Rozanska},
  M.~{Samorski}, H.~{Sander}, K.~{Sauerland}, C.~{Sese{\~n}a}, W.~{Stamm},
  {Methods to determine the angular resolution of the HEGRA extended air shower
  scintillator array}, Astroparticle Physics 5 (1996) 379--392.

\bibitem{Cobb:1999mi}
J.~H. Cobb, et~al., The observation of a shadow of the moon in the underground
  muon flux in the {Soudan~2} detector, Phys. Rev. D61 (2000) 092002.

\bibitem{Ambrosio:1998wv}
M.~Ambrosio, et~al., Observation of the shadowing of cosmic rays by the moon
  using a deep underground detector, Phys. Rev. D59 (1998) 012003.

\bibitem{Ambrosio:2003mz}
M.~Ambrosio, et~al., Moon and sun shadowing effect in the {MACRO} detector,
  Astropart. Phys. 20 (2003) 145--156.

\bibitem{Achard:2005az}
P.~Achard, et~al., Measurement of the shadowing of high-energy cosmic rays by
  the moon: A search for tev-energy antiprotons, Astropart. Phys. 23 (2005)
  411--434.

\bibitem{Bust}
Y.~M. {Andreyev}, V.~N. {Zakidyshev}, S.~N. {Karpov}, V.~N. {Khodov},
  {Observation of the Moon Shadow in Cosmic Ray Muons}, Cosmic Research 40
  (2002) 559--564.

\bibitem{Urban:1989eu}
M.~Urban, P.~Fleury, R.~Lestienne, F.~Plouin, Can we detect antimatter from
  other galaxies by the use of the earth's magnetic field and the moon as an
  absorber?, Nucl. Phys. Proc. Suppl. 14B (1990) 223--236.

\bibitem{Heintze:1990mq}
J.~Heintze, et~al., {Measuring the chemical composition of cosmic rays by
  utilizing the solar and geomagnetic fields}In Adelaide 1990, Proceedings,
  Cosmic ray, vol. 4 456- 459.

\bibitem{Parker:1958}
E.~N. Parker, Dynamics of interplanetary gas and magnetic fields, Astrophys. J.
  128 (1958) 664.

\bibitem{Wilcox:1965}
J.~M. {Wilcox}, N.~F. {Ness}, {Quasi-Stationary Corotating Structure in the
  Interplanetary Medium}, Journal of Geophysical Research 70 (1965) 5793--5805.

\bibitem{Amenomori:1993xj}
M.~Amenomori, et~al., {Direct evidence of the interplanetary magnetic field
  effect on the cosmic ray shadow by the sun}, Astrophys. J. 415 (1993)
  L147--L150.

\bibitem{SolarCycle}
{National Oceanographic and Atmospheric Administration},
  \url{http://www.swpc.noaa.gov/SolarCycle/}.

\bibitem{Adamson:2007ww}
P.~Adamson, et~al., Measurement of the atmospheric muon charge ratio at {TeV}
  energies with {MINOS}, Phys. Rev. D76 (2007) 052003.

\bibitem{IGRF}
I.~A. of~Geomagnetism \&~Aeronomy, International geomagnetic reference field
  IGRF-10.

\bibitem{Tsyganenko:1995}
N.~A. {Tsyganenko}, {Modeling the Earth's magnetospheric magnetic field
  confined within a realistic magnetopause}, Journal of Geophysical Research
  100 (1995) 5599--5612.

\bibitem{Tsyganenko:1996}
N.~A. {Tsyganenko}, D.~P. {Stern}, {Modeling the global magnetic field of the
  large-scale Birkeland current systems}, Journal of Geophysical Research 101
  (1996) 27187--27198.

\bibitem{Cash:1979vz}
W.~Cash, {Parameter estimation in astronomy through application of the
  likelihood ratio}, Astrophys. J. 228 (1979) 939--947.

\bibitem{CosB}
A.~M.~T. {Pollock}, et~al., {COS-B} gamma-ray sources and interstellar gas in
  the first galactic quadrant - caravane collaboration for the {COS-B}
  satellite, Astronomy and Astrophysics 146 (1985) 352--362.

\bibitem{Zhou:1997}
X.-W. {Zhou}, C.~T. {Russell}, G.~{Le}, N.~{Tsyganenko}, {Comparison of
  observed and model magnetic fields at high altitudes above the polar cap:
  POLAR initial results}, Geophysical Research Letters 24 (1997) 1451--1454.

\bibitem{omniweb}
Goddard space flight center, omniweb, \url{http://omniweb.gsfc.nasa.gov/}.

\end{thebibliography}
\end{document}